\documentclass[11pt]{article}

\usepackage[left=3cm,right=3cm,top=3cm,bottom=3cm]{geometry}
\usepackage{graphicx}
\usepackage{tikz}
\usetikzlibrary {arrows}
\usetikzlibrary {decorations}
\usepackage{natbib}

\usepackage{amssymb,amsmath,amsthm}

\newtheorem{theorem}{Theorem}

\newtheorem{definition}[theorem]{Definition}

 \newtheorem{algorithm}{Algorithm}
\numberwithin{equation}{section}
\allowdisplaybreaks

\title{Robust Bayesian
  Inference of Networks Using Dirichlet $t$-Distributions}

\author
{Michael Finegold\\
{mfinegol@andrew.cmu.edu} \\
Department of Statistics, Carnegie Mellon University, Pittsburgh, U.S.A. 
\and
Mathias Drton\\
{drton@uchicago.edu} \\
Department of Statistics, The University of Chicago, Chicago, U.S.A.}

\begin{document}

\maketitle

\begin{abstract}
  Bayesian graphical modeling provides an appealing way to obtain
  uncertainty estimates when inferring network structures, and much
  recent progress has been made for Gaussian models.  These models
  have been used extensively in applications to gene expression data,
  even in cases where there appears to be significant deviations from
  the Gaussian model. For more robust inferences, it is natural to
  consider extensions to $t$-distribution models.  We argue that the
  classical multivariate $t$-distribution, defined using a single
  latent Gamma random variable to rescale a Gaussian random vector, is
  of little use in highly multivariate settings, and propose other,
  more flexible $t$-distributions. Using an independent Gamma-divisor
  for each component of the random vector defines what we term the
  alternative $t$-distribution.  The associated model allows one to
  extract information from highly multivariate data even when most
  experiments contain outliers for some of their measurements.
  However, the use of this alternative model comes at increased
  computational cost and imposes constraints on the achievable
  correlation structures, raising the need for a compromise between
  the classical and alternative models.  To this end we propose the
  use of Dirichlet processes for adaptive clustering of the latent
  Gamma-scalars, each of which may then divide a group of latent
  Gaussian variables.  Dirichlet processes are commonly used to
  cluster independent observations; here they are used instead to
  cluster the dependent components of a single observation.  The
  resulting Dirichlet $t$-distribution interpolates naturally between
  the two extreme cases of the classical and alternative
  $t$-distributions and combines more appealing modeling of the
  multivariate dependence structure with favorable computational
  properties.\\
  
\end{abstract}

%%%%%%%%%%%%%%%%%%%%%%%%%%%%%%%%%%%%%%%%%%%%%%%%%%%

\section{Introduction}
\label{sec:intro}

Consider a random vector $\boldsymbol{Y}=(Y_1,\dots,Y_p)$ and an
undirected graph $G=(V,E)$ with vertex set $V=\{1,\dots,p\}$.  The
Gaussian graphical model given by the graph $G$ assumes that
$\boldsymbol{Y}$ follows a multivariate normal distribution
$\mathcal{N}_p(\boldsymbol{\mu},\boldsymbol{\Sigma})$, where
$\boldsymbol{\mu}$ may be any mean vector in $\mathbb{R}^p$ but
$\boldsymbol{\Sigma}$ is a positive definite covariance matrix that is
constrained such that two variables $Y_j$ and $Y_k$ are conditionally
independent given all the remaining variables
$\boldsymbol{Y}_{\setminus\{j,k\}}$ whenever $\{j,k\}$ is not an edge
in $E$.  The conditional independence holds if and only if
$\Sigma_{jk}^{-1}=0$; see e.g.~\cite{lauritzen}.  Hence, the graph
of a Gaussian graphical model can be inferred from data by
inferring the pattern of non-zero entries in the precision matrix
$\boldsymbol{\Sigma}^{-1}$.

Different Bayesian methods have been developed for
an uncertainty assessment in inference of the graph.
\cite{giudici}, for instance, use a uniform prior on decomposable
graphs and place a Hyper Inverse Wishart prior on the covariance
matrix,
which allows for exact local computation \citep{lauritzenHIW}.  In
particular, a closed form marginal likelihood permits treatment of
high-dimensional datasets \citep{dobra,carvalhostochastic}.  Exact
computations for non-decomposable graphs are much more involved
\citep{roveratond,dellaportas,atay}; for an approximate treatment see
\cite{lenkoski}.  Other recent literature providing different
extensions to the basic Gaussian model includes \cite{raj} and
\cite{carvalho}.

While appealing in many respects, Gaussian methods are susceptible to
great impact of measurement errors.  There is a substantial literature
on robustness in Bayesian inference, including \cite{definetti} and
\cite{westrobust}, but little work that directly targets graphical
models.  One commonly taken approach replaces multivariate normal by
multivariate $t$-distributions (with univariate $t$-margins).
$T$-distributions yield reasonable models for heavy-tailed marginal
behavior and have the capacity to maintain good statistical efficiency
when data are in fact Gaussian.  Moreover, convenient closed form
conditional distributions are available for use in Markov chain Monte
Carlo algorithms that exploit the representation of $t$-random
variables as ratios involving Gaussian and Gamma random variables.

The classical multivariate $t$-distribution can be defined in terms of
an unobserved Gaussian random vector and a single unobserved Gamma
divisor.  In the context of this paper, modeling assumptions then
refer to the latent Gaussian vector.
Penalized likelihood techniques for graphical model selection with
this classical $t$-distribution were explored in \cite{yuant}.  Highly
multivariate data, however, often have pockets of contamination in
many observations creating a scenario to which the classical
$t$-distribution is poorly suited.  This paper addresses this problem
by developing methods based on more flexible $t$-distributions.  The
distribution we term the alternative $t$-distribution has independent
Gamma scalars for each component of the latent Gaussian vector.  This
construction has been used in a frequentist treatment of graphical
models \citep{finegold}, but seems to have received little attention
otherwise.  While better suited to higher-dimensional analysis, the
distribution's use comes at increased computational cost and imposes
constraints on the achievable correlation structures; see Fig.~9 in
\cite{finegold}.  It is thus desirable to achieve a trade-off between
the two extremes, `classical' versus `alternative'.

The key new contribution of this paper is an adaptive method to
interpolate between the classical and the alternative case.  Our
proposal uses Dirichlet processes to cluster the Gamma divisors
associated with the components of the latent Gaussian vector. The
common Gamma-value associated with a cluster is then used to divide
all the associated Gaussian variables.  The Dirichlet process
clustering thus pertains to the possibly dependent components of a
single multivariate observation, as opposed to the common case of
clustering different independent observations.  Compared to the
alternative case, the Dirichlet $t$-proposal alleviates constraints on
correlations and reduces computational effort when a small number of
divisors is sufficient.  While we are concerned with graphical models,
there is no limitation to the use of the Dirichlet $t$-framework in
other problems of multivariate statistics.

The paper is organized as follows.  Section \ref{sec:armstrong} lays
out the Gaussian setup upon which our extensions are based.  In
Section \ref{sec:bayest}, we describe Bayesian inference of network
structures with classical and alternative $t$-distributions.  The
Dirichlet $t$-distribution is developed in Section
\ref{sec:dirichlettsection}.  Numerical experiments (Section
\ref{sec:simbayes}) and an analysis of gene expression data (Section
\ref{sec:bayesgasch}) demonstrate that our new framework is
computationally tractable and statistically efficient across a broad
spectrum of data with outliers.  A conclusion is given Section
\ref{sec:discussion}.

\section{A Bayesian Gaussian Graphical  Model}
\label{sec:armstrong}

\subsection{Background}
\label{subsec:background}

Let $IW_p(m,\boldsymbol{\Phi})$ denote the Inverse Wishart (IW)
distribution with degrees of freedom $m>p-1$ and a positive definite
scale matrix $\boldsymbol{\Phi}$.  This distribution is supported on
the cone of $p \times p$ positive definite matrices and has 
density
\begin{align}
  \label{eq:IW-density}
  p(\boldsymbol{\Sigma} \mid  m,\boldsymbol{\Phi}) &= \frac{|\frac{\boldsymbol{\Phi}}{2}|^{\frac{m}{2}}}{\Gamma_p(\frac{m}{2})}
|\boldsymbol{\Sigma}|^{-\frac{m+p+1}{2}} \exp\left\{-\frac{1}{2} \mathrm{tr}\left(\boldsymbol{\Phi}
    \boldsymbol{\Sigma}^{-1}\right)\right\}, 
\end{align}
where the determinant of a matrix $\boldsymbol{A}$ is denoted $|\boldsymbol{A}|$ and
\begin{align*}
\Gamma_p(\alpha) = \pi^{p(p-1)/4} \prod\limits_{i=1}^p \Gamma\left(\alpha -
  \left(\frac{i-1}{2}\right)\right) 
\end{align*}
is the multivariate gamma function with argument $\alpha > (p-1)/2$.
The distribution is the conjugate prior for the covariance matrix of a
multivariate normal distribution.  

Let $G=(V,E)$ be a decomposable graph with vertex set
$V=\{1,\dots,p\}$, and suppose $C_1, S_2,C_2,\dots,S_m,C_m$ is a
perfect sequence of the graph's cliques $C_i$ and separators $S_i$.
Here and throughout the paper, we assume familiarity with
basic graphical concepts as introduced in 
\cite{lauritzen}.  Let $M^+(G)$ be the set of positive definite
matrices $\boldsymbol{\Sigma}$ with entry $\Sigma_{jk}=0$ whenever
$\{j,k\}$ is not an edge in $E$.  For Gaussian graphical modeling, one
needs a constrained version of the IW distribution for the covariance
matrix $\boldsymbol{\Sigma}$ that has support such that the precision
matrix $\boldsymbol{\Sigma}^{-1}$ lies in $M^+(G)$.  The relevant
distribution is known as the Hyper IW distribution and we denote it by
$\mathit{HIW}(G,\delta,\boldsymbol{\Phi})$, where $\delta>0$ is a
degrees of freedom parameter and $\boldsymbol{\Phi}$ is a positive
definite scale matrix.  The distribution's  density
\begin{align*}
  f(\boldsymbol{\Sigma}\mid  G,\delta,\boldsymbol{\Phi})&= \frac{\prod\limits_{i=1}^m p(\boldsymbol{\Sigma}_{C_i
      C_i}\mid \delta+|C_i|-1,\boldsymbol{\Phi}_{C_i C_i})}{ \prod\limits_{i=2}^m
    p(\boldsymbol{\Sigma}_{S_iS_i}\mid \delta+|S_i|-1,\boldsymbol{\Phi}_{S_i S_i})}
\end{align*}
is the ratio of products of evaluations of the IW density from
(\ref{eq:IW-density}).  It follows from properties of the IW
distribution that the normalizing constant for the $\mathit{HIW}$
distribution is:
\begin{align}
  \label{eq:hiwnormal}
  h(G,\delta,\boldsymbol{\Phi}) &=  \frac{\prod\limits_{i=1}^m \left| \frac{\boldsymbol{\Phi}_{C_i C_i}}{2} \right|^{(\delta+|C_i|-1)/2} \Gamma_{|C_i|} \left( \frac{\delta+|C_i|-1}{2} \right)^{-1}}{\prod\limits_{i=2}^m \left| \frac{\boldsymbol{\Phi}_{S_i S_i}}{2} \right|^{(\delta+|S_i|-1)/2} \Gamma_{|S_i|} \left( \frac{\delta+|S_i|-1}{2} \right)^{-1}}.
\end{align}

\subsection{Model Specification}
\label{subsec:mode-specification}

We will treat a particular Bayesian Gaussian graphical model that is
similar to models in work such as \cite{armstrong}.  The considered
prior on graphs, $P(G)$, is supported on the set of decomposable
graphs with the probabilities of graphs proportional to
\begin{align}
  \label{eq:graph-prior-d}
  d^{|E|}(1-d)^{{p \choose 2}-|E|}.
\end{align}
The parameter $d$ controls the sparsity of the graph.  Conditional on
$G$ and hyperparameters $\delta$ and $\boldsymbol{\Phi}$, we let the
covariance matrix $\boldsymbol{\Sigma}$ follow a
$\mathit{HIW}(G,\delta,\boldsymbol{\Phi})$ distribution, which has the
consistency property that the submatrix distribution
$P(\boldsymbol{\Sigma}_{C_i C_i} \mid G,\delta,\boldsymbol{\Phi})$ is
the same for any graph $G$ with clique $C_i$ \citep{lauritzenHIW}.
Larger $\delta$ makes the posterior more concentrated around the
hyperparameter $\boldsymbol{\Phi}$.
As in \cite{carvalho} and \cite{donnet}, we choose $\delta=1$.
We use matrix hyperparameter $\boldsymbol{\Phi}=c
\boldsymbol{\mathcal{I}}_p$, a scalar multiple of the $p\times p$
identity matrix.  Larger $c$ leads to larger graphs; see
\cite{carvalhostochastic}.

Finally, suppose we observe a sample of $n$ independent
$\mathcal{N}_p(\boldsymbol{0},\boldsymbol{\Sigma})$ random vectors
$\boldsymbol{Y}_1,\dots, \boldsymbol{Y}_n$.  Let
$\boldsymbol{Y}$ be the matrix with the vectors $\boldsymbol{Y}_i$
as rows.  The joint distribution of
$(\boldsymbol{Y},G,\boldsymbol{\Sigma})$ then factors as
\begin{align*}
P(\boldsymbol{Y},G,\boldsymbol{\Sigma}\mid \delta,\boldsymbol{\Phi}) &= P(G) P(\boldsymbol{\Sigma}\mid
G,\delta,\boldsymbol{\Phi}) \prod\limits_{i=1}^n P(\boldsymbol{Y}_i\mid \boldsymbol{\Sigma}). 
\end{align*}
Centering Gaussian data by subtracting off the sample mean and
assuming mean $\boldsymbol{\mu}=\boldsymbol{0}$ is standard practice
since the distribution theory is essentially unchanged.

\subsection{Metropolis-Hastings Sampler}
\label{subsec:Metropolis}

We now briefly review the posterior sampling scheme used in prior work such
as \cite{armstrong}.  Define the sample covariance matrix
\begin{align*}
\boldsymbol{S} = (s_{jk}) &= \frac{1}{n} \sum_{i=1}^n \boldsymbol{\boldsymbol{Y}}_i\boldsymbol{Y}_i^T,
\end{align*}
and set $\boldsymbol{\Phi}^* = \boldsymbol{\Phi} +n\boldsymbol{S}$ and $\delta^* =\delta+n$.  Then
\begin{align}
\label{eq:hiwpost}
(\boldsymbol{\Sigma}\mid  \boldsymbol{Y},\delta,\boldsymbol{\Phi},G) &\sim \mathit{HIW}(G,\delta^*,\boldsymbol{\Phi}^*)
\end{align}
and 
\begin{align}
P(\boldsymbol{Y}\mid  \delta,\boldsymbol{\Phi},G) &= 
\frac{1}{ (2 \pi)^{np/2}}\cdot
\frac{h(G,\delta,\boldsymbol{\Phi})}{h(G,\delta^*,\boldsymbol{\Phi}^*)}; 
\end{align}
see \cite{lauritzenHIW} and \cite{giudici}.  

Let $G=(V,E)$ and $G'=(V,E')$ be two decomposable graphs on
$V=\{1,\dots,p\}$.  Suppose that $C_1,S_2,C_2,\dots,S_m,C_m$ is a perfect
sequence of the cliques and separators of $G$ and that $\{j,k\} \in E$.  If
$G'$ is equal to $G$ except that the edge $\{j,k\}$ is removed, then the
following three properties hold; see e.g.~\cite{armstrong}.
First, the edge $\{j,k\}$ is in a single clique $C_q$ of $G$.  Second, we
have either $j \notin S_q$ or $k \notin S_q$.  Third, suppose $k \notin
S_q$, and let $C_{q_1}=C_q \setminus \{k\}$, $C_{q_2}=C_q \setminus \{j\}$
and $S_{q_2}=C_q \setminus \{j,k\}$. Then
$C_1,S_2,\dots,S_q,C_{q_1},S_{q_2},C_{q_2},S_{q+1},\dots,S_m,C_m$ is a
perfect ordering of all cliques and separators of $G'$.  Let
$\delta_2=\delta+|S_{q_2}|$ and $\delta_2^*=\delta^*+|S_{q_2}|$.  The above
three observations can be used to show that the ratio of marginal
likelihoods $P(\boldsymbol{Y}\mid G)/P(\boldsymbol{Y}\mid G')$ is equal to
\begin{align}
  \label{eq:MH}
  \frac{h(G,\delta,\boldsymbol{\Phi})h(G^{\prime},\delta^*,\boldsymbol{\Phi}^*)}{h(G,\delta^*,\boldsymbol{\Phi}^*)h(G^{\prime},\delta,\boldsymbol{\Phi})} 
  &= 
  \frac{ |\boldsymbol{\Phi}_{ee|S_{q_2}}|^{\left(\frac{\delta_2 +1}{2} \right)}  
    \left|\boldsymbol{\Phi}^*_{jj|S_{q_2}}
      \boldsymbol{\Phi}^*_{kk|S_{q_2}}\right|^{\left(\frac{\delta^*_2}{2} \right)}}{ 
    |\boldsymbol{\Phi}^*_{ee|S_{q_2}}|^{\left(\frac{\delta^*_2+1}{2} \right)}
    \left|\boldsymbol{\Phi}_{jj|S_{q_2}}  
      \boldsymbol{\Phi}_{kk|S_{q_2}}\right|^{\left(\frac{\delta_2}{2} \right)}  
  }  \times
  \frac{ \Gamma \left( \frac{\delta_2}{2} \right)
    \Gamma \left( \frac{\delta^*_2+1}{2} \right)}{
    \Gamma \left( \frac{\delta_2+1}{2} \right)\Gamma \left( \frac{\delta^*_2}{2} \right)}.
\end{align}
Here, $e=\{j,k\}$ and $\boldsymbol{\Phi}_{ee|S_{q_2}} =
\boldsymbol{\Phi}_{ee} - \boldsymbol{\Phi}_{eS_{q2}}
(\boldsymbol{\Phi}_{S_{q2}S_{q2}})^{-1} \boldsymbol{\Phi}_{S_{q2}e}$;
the conditional variances for $j$ and $k$ are defined similarly.  The
ratio in (\ref{eq:MH}) allows one to create a Markov chain with the
posterior distribution $P(G\mid \boldsymbol{Y})$ as the stationary
distribution by applying a Metropolis-Hastings procedure that avoids
sampling of $\boldsymbol{\Sigma}$.

\begin{algorithm}[Gaussian]
  \label{MHprocedure}
  Starting with a decomposable graph $G_0$, repeat the following
  two steps for $t=0,1,2,\dots$: \\
  (i) Create a graph $G'$ by randomly picking an edge to delete from
  $G_t$ or  to add to $G_t$.\\
  (ii) If $G'$ is decomposable, accept the move $G_{t+1}=G'$ with
  probability
  \begin{align}
    \min \left\{1,\frac{P(\boldsymbol{Y}\mid G^ {\prime})}{P(\boldsymbol{Y}\mid G)} \right\},
  \end{align}
  setting $G_{t+1}=G$ if the move is rejected or $G'$ is not decomposable.
\end{algorithm}
 
Decomposability of the input $G_0$ can be tested with the
Max-Cardinality algorithm \citep{lauritzenexpert}.  Given the
decomposable graph $G_0$, the set of all decomposable graphs can be
traversed following simple rules for edge addition and deletion
\citep{giudici}.

\section{Bayesian Graphical Models for $t$-Distributions}
\label{sec:bayest}

\subsection{Classical and Alternative Multivariate $t$-Distributions}
\label{sec:t-dist}

The {\em classical\/} multivariate $t$-distribution
$t_{p,\nu}(\boldsymbol{\mu},\boldsymbol{\Psi})$ in $\mathbb{R}^p$ has
density
\begin{equation}
  \label{eq:tdensity}
  f_\nu(\boldsymbol{y}\mid  \boldsymbol{\mu},\boldsymbol{\Psi}) = \frac{\Gamma (\frac{\nu+p}{2}) |\boldsymbol{\Psi}|^{-1/2}}{(\pi
  \nu )^{p/2} \Gamma 
  (\frac{\nu}{2}) [1 + \delta_y (\boldsymbol{\mu}, \boldsymbol{\Psi})/\nu]^{(\nu+p)/2}} ,
\end{equation}
where $\delta_y(\boldsymbol{\mu},\boldsymbol{\Psi}) =
(\boldsymbol{y}-\boldsymbol{\mu})^T \boldsymbol{\Psi}^{-1}
(\boldsymbol{y}-\boldsymbol{\mu})$ and $\boldsymbol{y}
\in\mathbb{R}^p$.  The vector $\boldsymbol{\mu}\in\mathbb{R}^p$ is the
mean vector.  The scale parameter matrix
$\boldsymbol{\Psi}=(\psi_{jk})$ is assumed positive definite.  For
degrees of freedom $\nu>2$, the covariance matrix exists and is equal
to $\nu/(\nu-2)$ times $\boldsymbol{\Psi}$.
If $\boldsymbol{X}\sim\mathcal{N}_p(\boldsymbol{0},\boldsymbol{\Psi})$
is a multivariate normal random vector independent of the Gamma-random
variable $\tau\sim\Gamma(\nu/2,\nu/2)$, then
$\boldsymbol{Y}=\boldsymbol{\mu}+\boldsymbol{X}/\sqrt{\tau}$ has a
$t_{p,\nu}(\boldsymbol{\mu},\boldsymbol{\Psi})$-distribution
\citep[Chap.~1]{kotz}.  The heavy tails of the distribution are
related to small values of the divisor $\tau$.

The classical $t$-distribution is useful for robust inference when only a
handful of the observations are contaminated.
When the dimension $p$ is large, however, it is not uncommon for
contamination to affect rather small parts of many observations.  To
handle such a situation better we consider $p$ independent
divisors $\tau_1,\dots,\tau_p \sim \Gamma(\nu/2,\nu/2)$.  Assuming the
divisors to be also independent of  $\boldsymbol{X}
\sim \mathcal{N}_p(0,\boldsymbol{\Psi})$, we create the random vector
$\boldsymbol{Y}$ with coordinates $Y_j = \mu_j + X_j/\sqrt{\tau_j}$
and define the {\em alternative\/} multivariate $t$-distribution to be
the joint distribution of $\boldsymbol{Y}$. In symbols,
$\boldsymbol{Y} \sim t^*_{p,\nu}(\boldsymbol{\mu},\boldsymbol{\Psi})$.
For robustified inference, the alternative $t$-distribution is
appealing as it allows different rescaling of the different components
of $\boldsymbol{Y}$.

\subsection{Bayesian Inference With Classical $t$-Distributions}
\label{sec:bayesclassical}
 
Suppose $\boldsymbol{Y}_1,\dots, \boldsymbol{Y}_n\in\mathbb{R}^p$ are
independent random vectors drawn from the classical multivariate
$t$-distribution $t_{\nu,p}(\boldsymbol{\mu},\boldsymbol{\Psi})$.  Let
$\boldsymbol{Y}$ be the matrix with the vectors $\boldsymbol{Y}_i$ as
rows.  We are interested in the posterior distribution on graphs,
$P(G\mid \boldsymbol{Y})$, where the graph $G$ corresponds to
conditional independence constraints on the latent multivariate normal
vectors $\boldsymbol{X}_i$, that is, an off-diagonal entry
$\theta_{jk}$ of the matrix
$\boldsymbol{\Theta}=\boldsymbol{\Psi}^{-1}$ is zero unless $\{j,k\}$
is an edge in $G$.

In the normal model we can center the data by subtracting off the sample
mean and assume, without loss of generality, that
$\boldsymbol{\mu}=\boldsymbol{0}$.  For the $t$-model, robust estimation of
both $\boldsymbol{\mu}$ and $\boldsymbol{\Psi}$ is desirable, and we thus
include $\boldsymbol{\mu}$ in our setup.  Let $\boldsymbol{\tau} = (\tau_1,
\dots , \tau_n)$ be the vector of unobserved Gamma-divisors for the $n$
observations $\boldsymbol{Y}_i$.  Proceeding as in the normal case, our
full model factors the joint distribution of $\boldsymbol{Y},
\boldsymbol{\tau}, G, \boldsymbol{\Psi},\boldsymbol{\mu}$ as
\begin{align}
\label{eq:bayesmodelwithmu}
P(\boldsymbol{Y},\boldsymbol{\tau},G,\boldsymbol{\Psi},\boldsymbol{\mu}) &= P(G)P(\boldsymbol{\mu})P(\boldsymbol{\Psi}\mid G,\delta,\boldsymbol{\Phi}) \prod\limits_{i=1}^nP(\boldsymbol{Y}_i\mid \tau_i,\boldsymbol{\Psi},\boldsymbol{\mu})P(\tau_i\mid \nu),
\end{align}
where
\begin{align}
(\boldsymbol{Y}_i\mid \tau_i,\boldsymbol{\Psi}) &\sim \mathcal{N}_p(\boldsymbol{\mu},\boldsymbol{\Psi}/\tau_i), &
\label{eq:condnormal}
(\boldsymbol{\Psi}\mid G,\delta,\boldsymbol{\Phi}) &\sim \mathit{HIW}(G,\delta,\boldsymbol{\Phi}), \\
 \label{eq:psihiw}
(\tau_i\mid \nu) &\sim \Gamma(\nu/2,\nu/2), 
&
\boldsymbol{\mu} &\sim \mathcal{N}_p(\boldsymbol{0},\sigma_{\mu} \cdot \boldsymbol{\mathcal{I}}_p).
\end{align}
In later numerical work, we choose $\sigma_{\mu}$ large enough for the
prior to be ``flat'' over a range of plausible values.
Throughout the hyperparameters $\delta$, $\boldsymbol{\Phi}$, and $\nu$ are fixed; recall Section~\ref{subsec:mode-specification}.

Inference for the Gaussian case is simplified by integrating out the
covariance matrix $\boldsymbol{\Sigma}$; recall~(\ref{eq:MH}).
In the $t$-model, we may condition on $\tau$ and add/remove
edges based on the ratio
\begin{align}
\label{eq:MHratiot}
\frac{P(\boldsymbol{Y}\mid G^ {\prime},\tau,\boldsymbol{\mu})}{P(\boldsymbol{Y}\mid G,\tau,\boldsymbol{\mu})} &=
\frac{h(G,\delta,\boldsymbol{\Phi})h(G^{\prime},\delta^*,\boldsymbol{\Phi}^*_{\tau})}{h(G,\delta^*,\boldsymbol{\Phi}^*_{\tau})h(G^{\prime},\delta,\boldsymbol{\Phi})}, 
\end{align}
where
\begin{align}
\boldsymbol{S}_{\boldsymbol{\tau} \boldsymbol{Y}\boldsymbol{Y}}(\boldsymbol{\mu}) &= \frac{1}{n} \sum_{i=1}^n \tau_i (\boldsymbol{Y}_i -
  \boldsymbol{\mu})(\boldsymbol{Y}_i- \boldsymbol{\mu})^T&
  & \mathrm{ and } &
\boldsymbol{\Phi}^*_{\tau} &= \boldsymbol{\Phi} + n\boldsymbol{S}_{\boldsymbol{\tau} \boldsymbol{Y}\boldsymbol{Y}}(\boldsymbol{\mu}).
\end{align}
Drawing $\boldsymbol{\tau}$ given $G,\boldsymbol{\mu}$ and
$\boldsymbol{Y}$ is difficult, however, and we resort to conditioning
on further parameters and consider the conditional distribution
\begin{equation}
\label{eq:tau-given-Y}
  (\tau_i \mid  \boldsymbol{Y},\boldsymbol{\mu},\boldsymbol{\Psi}) \sim \Gamma\left( \frac{\nu + p}{2}, \frac{\nu +
    \delta_{\boldsymbol{Y}_i}(\boldsymbol{\mu},\boldsymbol{\Psi})}{2}\right);
\end{equation}
compare \cite{liurubin}.  This requires
$\boldsymbol{\Theta}=\boldsymbol{\Psi}^{-1}$, and we use the method of
\cite{carvalhosim} to draw from
\begin{align}
  \label{eq:hiwpost-Psi}
  (\boldsymbol{\Psi}\mid \boldsymbol{Y},G,\tau,\boldsymbol{\mu})&\sim \mathit{HIW}(G,\delta^*,\boldsymbol{\Phi}^*_{\tau}).
\end{align}
%From \cite{roverato} we then know the posterior of $\boldsymbol{\Theta}\mid G,Y,\tau$ is
%\begin{align}
%P(\boldsymbol{\Theta}\mid G,Y,\tau,\boldsymbol{\mu}) = C \cdot |\boldsymbol{\Theta}|^{(\delta^*-2)/2} \exp \{ - \frac{1}{2} tr(\boldsymbol{\Theta} \boldsymbol{\Phi}_{\tau}^*) \}
%\end{align}
%for positive definite $\boldsymbol{\Theta}$.  We could then propose a new $\boldsymbol{\Theta}$, check if it is positive definite, and then accept the proposal using a Metropolis ratio based on the above posterior.  
%Alternatively, 
This procedure first draws $\boldsymbol{\Psi}_{C_1C_1}$, cycles through the
cliques of $G$ to draw $\boldsymbol{\Psi}_{C_iC_i}$ given
$\boldsymbol{\Psi}_{S_i,S_i}$, and then uses a standard completion
algorithm to determine the values of $\boldsymbol{\Psi}$ not associated
with any clique.  We then invert $\boldsymbol{\Psi}$ to obtain
$\boldsymbol{\Theta}$.  For decomposable graphs, the inversion can be done
efficiently using the procedure of \cite{lauritzenHIW}.  That is, we
compute
\begin{align}
\label{eq:invtheta}
\boldsymbol{\Theta} &= \sum\limits_{i=1}^m {(\boldsymbol{\Psi}_{C_i C_i})^{-1}}^{[0]} -  
\sum\limits_{i=2}^m {(\boldsymbol{\Psi}_{S_i S_i})^{-1}}^{[0]} ,
\end{align}
where ${(\boldsymbol{\Psi}_{C_i C_i})^{-1}}^{[0]}$ means that we take
the $p \times p$ matrix of zeros and add in the elements of
${(\boldsymbol{\Psi}_{C_i C_i})^{-1}}$ in their appropriate places.
This calculation only requires the elements $\psi_{jk}$ of
$\boldsymbol{\Psi}$ corresponding to edges $\{j,k\} \in E$.
Therefore, for the purposes of obtaining $\boldsymbol{\Theta}$, we
need not perform the completion step in the method of
\cite{carvalhosim}.  Moreover, every step in the generation and
inversion of $\boldsymbol{\Psi}$ is based on local computations at the
clique level.

Now the conditional distribution $(\boldsymbol{\mu}\mid
\boldsymbol{Y},\boldsymbol{\tau},\boldsymbol{\Theta},G)$ is the
multivariate normal distribution
\begin{align}
\label{eq:mupost}
\mathcal{N}_p\left(\left[\bigg(\sum_{i=1}^n\tau_i\bigg)\boldsymbol{\Theta}
    +
    \boldsymbol{\Theta}_{\mu}\right]^{-1}\boldsymbol{\Theta}\bigg(\sum_{i=1}^n
    \tau_i \boldsymbol{Y}_i 
      \bigg),\left[\bigg(\sum_{i=1}^n\tau_i\bigg)\boldsymbol{\Theta}
    + \boldsymbol{\Theta}_{\mu}\right]^{-1} \right)
\end{align}
where $\boldsymbol{\Theta}_{\mu} =
\boldsymbol{\mathcal{I}}_p/\sigma_{\mu}$.  To draw $\boldsymbol{\mu}$
using this conditional distribution we must invert the $p\times p$
matrix $\big(\sum_{i=1}^n\tau_i\big)\boldsymbol{\Theta} +
\boldsymbol{\Theta}_{\mu}$, a potentially computationally expensive
procedure.  For practical applications, we thus simply set
    \begin{align}
    \label{eq:approxmu}
    \boldsymbol{\mu} =  \frac{\sum_{i=1}^n \tau_i \boldsymbol{Y}_i}{\sum_{i=1}^n
  \tau_i}
  \end{align}
  instead of drawing from the conditional distribution in 
  \eqref{eq:mupost}.  We provide theoretical and numerical
  justifications for this alternative in the Appendix.

\begin{algorithm}[Classical $t$]
  \label{alg:gibbsclassical}
  Starting with a decomposable graph $G_0$, and initial values $\boldsymbol{\mu}_0$ and
  $\boldsymbol{\tau}_0$, repeat the following steps for $t=0,1,2,\dots$: \\ 
  (i) Jointly draw a new graph $G_{t+1}$ and a new matrix $\boldsymbol{\Theta}_{t+1}$ as
  follows:
  \begin{enumerate}
  \item[(a)]  Draw $G_{t+1}$ as in Algorithm \ref{MHprocedure}, but using
  the ratio in (\ref{eq:MHratiot}). 
\item[(b)] Conditional on $(\boldsymbol{Y},G_{t+1},\boldsymbol{\tau}_t,\boldsymbol{\mu}_t)$, sample $\boldsymbol{\Theta}_{t+1}$
  by drawing $\boldsymbol{\Psi}_{t+1}$ from (\ref{eq:hiwpost-Psi}) and inverting it using
  (\ref{eq:invtheta}).
  \end{enumerate}
  (ii) Conditional on $(\boldsymbol{Y},G_{t+1},\boldsymbol{\Theta}_{t+1})$, sample the new
  independent components of the vector
  $\boldsymbol{\tau}_{t+1}=(\tau_{t+1,1},\dots,\tau_{t+1,n})$ from
  (\ref{eq:tau-given-Y}).\\
  (iii) Set $\boldsymbol{\mu}_{t+1}$ to the value in (\ref{eq:approxmu}).
\end{algorithm}

In practice we hope to improve on the estimate of $P(G\mid
\boldsymbol{Y})$ that we would obtain from the normal model.  If we
start with a ``good'' estimate of $\boldsymbol{\tau}$ as given by the
{\em tlasso} of \cite{finegold}, we may be able to make considerable
improvement over the normal model without sampling $\boldsymbol{\tau}$
after every edge draw.  In our later simulations, we thus draw
$\boldsymbol{\tau}$ only every $k>1$ draws of
$(G,\boldsymbol{\Theta})$, which does not affect the validity of the
sampler.  Note also that the $k-1$ intermediate iterations do not
involve $\boldsymbol{\Psi}$ (or its inverse $\boldsymbol{\Theta}$).

\subsection{Bayesian Inference With Alternative $t$-Distributions}
\label{sec:altgibbs}

For the alternative $t$-model, there are only a few differences.  The model
itself is the same as in (\ref{eq:bayesmodelwithmu}) except that now
$\boldsymbol{\tau}_i$ is a $p$-vector, 
 and
\begin{align}
\label{eq:bayesaltmodel}
(\boldsymbol{Y}_i\mid
\boldsymbol{\tau}_i,\boldsymbol{\Psi},\boldsymbol{\mu}) &\sim
\mathcal{N}_p(\boldsymbol{\mu},\mathrm{diag}(1/\sqrt{\boldsymbol{\tau}_i})\cdot
\boldsymbol{\Psi} \cdot \mathrm{diag}(1/\sqrt{\boldsymbol{\tau}_i})),\nonumber \\
(\tau_{ij}\mid \nu) &\sim \Gamma(\nu/2,\nu/2), \quad j=1,\dots,p,
\nonumber
\end{align}
with $\tau_{i1},\dots,\tau_{ip}$ independent given $\nu$.  For the
Gibbs sampler, we cannot draw the matrix
$\boldsymbol{\tau}=(\tau_{ij})$ given
$\boldsymbol{Y},\boldsymbol{\Theta},G,\boldsymbol{\mu}$ directly, but
we can draw $\tau_{ij}$ given $\boldsymbol{\tau}_{i \setminus
  j},\boldsymbol{Y},\boldsymbol{\Theta},G,\boldsymbol{\mu}$.  The
conditional
density (derived in the Appendix) is
\begin{equation}
  \label{eq:full-conditional}
  f(\tau_{ij}\mid \boldsymbol{\tau}_{ i \setminus j},\boldsymbol{Y}) = C(\alpha,\beta,\gamma)\cdot
  \tau_{ij}^{\alpha 
  - 1} \exp \left\{-\tau_{ij} \beta - \sqrt{\tau_{ij}} \gamma \right\} 
\end{equation}
with
\begin{align*}
\alpha &= \frac{\nu+1}{2}, 
& \beta &= \frac{\nu + (Y_{ij}-\mu_j)^2\theta_{jj}}{2}, 
&
\gamma &= (Y_{ij}-\mu_j)\boldsymbol{\Theta}_{j \setminus j}\boldsymbol{X}_{i \setminus j},
\end{align*}
and normalizing constant $C(\alpha,\beta,\gamma)$.  The problem of
sampling from this density also arose in the work of \cite{finegold},
where the density appears in the context of a Markov chain Monte Carlo
EM algorithm.  We employ a rejection sampling scheme from
\cite{ying:2012}.  This leads to the following Gibbs sampler.

\begin{algorithm}[Alternative $t$]
  \label{alg:gibbsalternative}
  Starting with a decomposable graph $G_0$, and initial values
  $\boldsymbol{\mu}_0$ and
  $\boldsymbol{\tau}_0$, repeat the following steps for $t=0,1,2,\dots$: \\
  (i) Jointly draw a new graph $G_{t+1}$ and a new matrix
  $\boldsymbol{\Theta}_{t+1}$ as
  in Algorithm \ref{alg:gibbsclassical}.\\
  (ii) For each observation $i=1,\dots,n$, cycle through the variables
  $j=1,\dots,p$ and draw $\tau_{ij}$ from its current full conditional
  in (\ref{eq:full-conditional}) to obtain
  a new matrix $\boldsymbol{\tau}_{t+1}$.\\
  (iii) Set $\boldsymbol{\mu}_{t+1}$ to the value in
  (\ref{eq:approxmu}), where $\boldsymbol{\tau}$ is now a vector and
  the multiplications and divisions are done component-wise.
\end{algorithm}
 
This sampling scheme for the alternative model works well for moderate
$p$ ($p\approx 100$) and underlies our later simulations.  The scheme
becomes very computationally intensive, however, for large $p$, both
in terms of the time to complete one iteration of step (ii) above and
the number of iterations required to approach convergence of the
Markov chain.  It is conceivable that other strategies, such as using
a Metropolis-Hastings step to sample from
$P(\boldsymbol{\tau}|G,\boldsymbol{Y})$ directly, might perform
better.  
However, we will not treat such alternative sampling schemes in the
remainder of this paper, which is instead devoted to other models.

\section{Dirichlet $t$-Models}
\label{sec:dirichlettsection}

We are faced with a trade-off between the classical and alternative
models.  If our goal is to identify pockets of contamination spread
throughout a large data set, we certainly do not want to weight an
entire observation via a single divisor as in the classical model.
In the other extreme, with a different divisor for each variable, the
alternative model proves to be computationally burdensome.  Moreover,
the alternative model has pairwise correlations bounded at a level
that is somewhat restrictive for small degrees of freedom $\nu$; see
\cite{finegold}.  The approach we propose in this section interpolates
between the two extremes and seems appealing in particular when there
are batches of variables taking on extreme values, while the rest
exhibit behavior consistent with a normal model.

If groups of variables are similarly contaminated (or otherwise
extreme), we can share statistical strength and ease our computational
burden by forming clusters of Gamma divisors for each observation.  We
solve this clustering problem via a Dirichlet Process (DP) prior on
the vector of $\boldsymbol{\tau}$ divisors for each observation.  This
Bayesian nonparametric approach avoids fixing a number of clusters and
truly interpolates between the classical and alternative case.

\subsection{Background on Dirichlet Processes}

The Dirichlet Process is a measure on measures introduced by
\cite{ferguson}.  Let $P_0$ be a probability measure on a measurable
space $(\boldsymbol{\Theta},\mathcal{B})$, and $\alpha>0$.  We say
that $P$ is distributed according to a Dirichlet process with
parameters $\alpha$ and $P_0$ if for any finite measurable partition
$(A_1,\dots,A_r)$ of $\boldsymbol{\Theta}$, the random vector
$(P(A_1),\dots,P(A_r))$ follows a Dirichlet distribution with
parameters $(\alpha P_0(A_1),\dots,\alpha P_0(A_r))$.  We write $P
\sim DP(\alpha,P_0)$.

The Dirichlet process possesses a clustering property due to the fact
that if $P \sim DP(\alpha,P_0)$ then $P$ is discrete with probability
$1$.  This holds even if the base measure $P_0$ is continuous
\citep{ferguson}.  Let $\pi_1,\dots,\pi_n$ be independent draws from a
random measure $P\sim DP(\alpha,P_0)$.  Then
the conditional distribution of $\pi_n$ given $\pi_{\setminus n}$ is a
mixture, namely,
\begin{align}
  \label{eq:rich-get-richer}
  (\pi_n \mid  \pi_{\setminus n}) &\sim \frac{\alpha}{\alpha + n-1}
  P_0(\pi_n) + \frac{n-1}{\alpha + n-1} \sum\limits_{j =1}^{n-1}
  \delta_{\pi_j}(\pi_n), 
\end{align}
where $\delta_{\pi_j}$ denotes a point mass at $\pi_j$.  Hence, each
new draw has a positive probability of assuming the same value as a
previous draw, and this probability increases with each new draw.  The
choice of $\alpha$ greatly influences the number of expected clusters,
with larger values leading to more clusters.  New observations taking
on the same values as existing ones in (\ref{eq:rich-get-richer})
gives an intuitive explanation to the phenomenon that Dirichlet
processes often produce a small number of large clusters.  This can be
unsuitable for generic clustering applications but is, in fact,
appealing for the robustification problem we consider.  Here, we might
often expect one large cluster that corresponds to uncontaminated
(high-quality) observations.

\subsection{Dirichlet $t$-Model}
\label{sec:dirichlett}

\begin{figure}
\centering
\begin{tikzpicture}[scale=1]
\tikzstyle{every node}=[draw,shape=circle];
\path (3.5,0) node (P0) {$P_0$};
\path (5.5,0) node (P) {$P$};
\path (7.5,0) node (a) {$\alpha$};

\path (2.5,2) node (t1) {$\tau_1$};
\path (5.5,2) node (t2) {$\tau_2$};
\path (8.5,2) node (t3) {$\tau_3$};

\path (3,4) node (X1) {$X_1$};
\path (6,4) node (X2) {$X_2$};
\path (9,4) node (X3) {$X_3$};

\path (4,2) node [shade] (Y1) {$Y_1$};
\path (7,2) node [shade] (Y2) {$Y_2$};
\path (10,2) node [shade] (Y3) {$Y_3$};

\draw (X1) -- (X2) (X2) -- (X3);
\draw[->] (P0) -- (P);
\draw[->] (a) -- (P);
\draw[->] (P) -- (t1);
\draw[->] (P) -- (t2);
\draw[->] (P) -- (t3);
\draw[->] (t1) -- (Y1);
\draw[->] (t2) -- (Y2);
\draw[->] (t3) -- (Y3);
\draw[->] (X1) -- (Y1);
\draw[->] (X2) -- (Y2);
\draw[->] (X3) -- (Y3);
\end{tikzpicture}
\caption{Representation of the process generating a $t^{\alpha}$-random vector
  $\boldsymbol{Y}$ from a latent normal random vector $\boldsymbol{X}$ and an independent Dirichlet Gamma random vector $\boldsymbol{\tau}$.  The missing undirected edge between $X_1$ and $X_3$ indicates a graphical conditional independence.  Directed arrows illustrate the functional
  relationship among $\boldsymbol{X}$, $\boldsymbol{\tau}$, and $\boldsymbol{Y}$.}
  \label{fig:dirichlet} 
\end{figure}
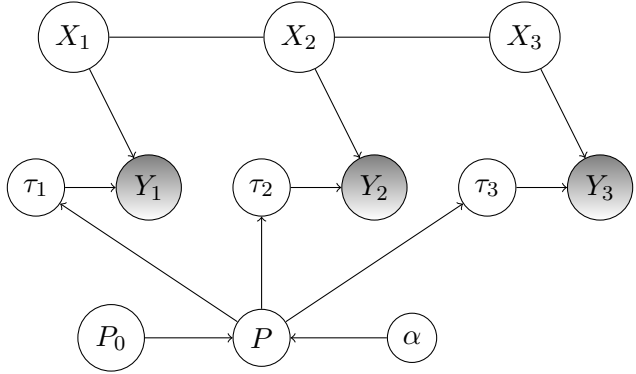

Applying Dirichlet processes in the $t$-distribution context yields
the following construction, illustrated in Figure~\ref{fig:dirichlet}.

\begin{definition}
\label{def:dirichlett}
Let $P_0$ be the $\Gamma(\nu/2,\nu/2)$ distribution and let $P \sim
DP(\alpha,P_0)$.  For $j=1, \ldots,p$, let $\tau_j \sim P$ be
independent of each other given $P$.  We then say that the random
vector $\boldsymbol{\tau}\in\mathbb{R}^p$ follows a Dirichlet Gamma
distribution; in symbols $\boldsymbol{\tau} \sim
D\Gamma_p(\alpha,\nu)$.  If the random vectors $\boldsymbol{X} \sim
\mathcal{N}_p(\boldsymbol{0},\boldsymbol{\Psi})$ and
$\boldsymbol{\tau} \sim D\Gamma_p(\alpha,\nu)$ are independent, then
we say that  $\boldsymbol{Y}\in\mathbb{R}^p$ with
coordinates $Y_j = \mu_j +X_j/\sqrt{\tau_j}$ follows a Dirichlet
$t$-distribution; in symbols $\boldsymbol{Y} \sim
t^{\alpha}_{p,\nu}(\boldsymbol{\mu},\boldsymbol{\Psi})$.
\end{definition}

The family of Dirichlet $t$-distribution includes the previous two models
as extreme cases.  When $\alpha \rightarrow 0$, we will have one cluster,
giving us the classical $t$-distribution.  When $\alpha \rightarrow \infty$
we will have $p$ clusters, giving us the alternative $t$-distribution.

Unlike in the alternative model, there is no upper bound on the correlations between two variables {\em within} a cluster.  For {\em any} two variables $Y_j$ and $Y_k$, the marginal covariance is 
\begin{align}
\psi_{jk} \left[ \frac{1}{\alpha+1} \frac{\nu}{\nu -2}  + \frac{\alpha}{\alpha+1} \frac{\nu \Gamma((\nu - 1)/2)^2}{2 \Gamma(\nu/2)^2} \right]
\end{align}
where $1/(\alpha+1)$ is the probability that any two variables will be
in the same $\tau$ cluster.  As $\alpha \rightarrow 0$ we obtain a
maximum covariance of $\psi_{jk} \cdot \nu/(\nu-2)$ and a maximum
correlation of 1.

In contrast to the mixture model of \cite{antoniak}, which uses $n$
draws from a Dirichlet process for $n$ observations, we here draw $p$
values for the coordinates of a single observation.  The relevant
conditional distributions are similar to those given in \cite{escobar}
but include terms that reflect that dependence among the $p$
coordinates of an observation.

For Gibbs sampling we need the following full conditional (derived
in the Appendix). 
For notational simplicity we consider $\boldsymbol{Y}$ and
$\boldsymbol{\tau}$ to be $p$-vectors representing a single
observation:
\begin{align}
\label{eq:clustering}
(\tau_j\mid \boldsymbol{\tau}_{\setminus j},\boldsymbol{Y},\boldsymbol{\Theta}) &\sim q_0 P_j(\tau_j) + \sum\limits_{j^{\prime} \ne j} q_{j^{\prime}} \delta_{\tau_{j^{\prime}}}(\tau_j).
\end{align}
This mixture distribution involves point masses, denoted
$\delta_{\tau_k}(\tau_j)$ when supported at $\tau_k$, and the distribution
$P_j(\tau_j)$, which is the conditional distribution of $\tau_j$ given
$(\boldsymbol{\tau}_{\setminus j},\boldsymbol{Y},\boldsymbol{\Theta})$ from
the alternative $t$-model.  The mixture weights for the point masses are
denoted $q_{j^{\prime}}$ and are proportional to the conditional density of
$(Y_j\mid \boldsymbol{\Theta},\boldsymbol{Y}_{\setminus
  j},\boldsymbol{\tau}_{\setminus j},\tau_j=\tau_{j^{\prime}})$ evaluated
at $y_j$.  This density is that of a normal distribution with mean
$\mu_c/\sqrt{\tau_{j^{\prime}}}$ and variance
$\sigma_c^2/\tau_{j^{\prime}}$ evaluated at $y_j-\mu_j$.  We denote the
density as
$f_{\mathcal{N}}(y_j-\mu_j;\mu_c/\sqrt{\tau_{j^{\prime}}},\sigma^2_c/\tau_{j^{\prime}})$.
Finally, the remaining mixture weight $q_0$ is proportional to $\alpha$
times the conditional density of $(Y_j\mid
\boldsymbol{\Theta},\boldsymbol{Y}_{\setminus j},
\boldsymbol{\tau}_{\setminus_j})$ evaluated at $y_j$, where $\tau_j\sim
P_0$.  This density is that of a noncentral $t$-distribution with degrees
of freedom $\nu$ and noncentrality parameter $\mu_c/\sigma_c$, where
$\mu_c$ and $\sigma_c$ are the conditional mean and standard deviation of
$(X_j\mid \boldsymbol{X}_{\setminus j})$.  We denote the density as
$f_T(y_j-\mu_j;\mu_c/\sigma_c,\nu)/\sigma_c$.

Now define a vector $\boldsymbol{z}\in\mathbb{R}^p$ of cluster
indicators by setting $z_j=k$ if $\tau_j$ belongs to the $k^{th}$
cluster.  In our setting, all $\tau_j$ in the $k^{th}$ cluster assume
the same value, $\eta_k$, and thus $\boldsymbol{\tau}$ is a function of
$\boldsymbol{z}$ and the vector of cluster values $\boldsymbol{\eta}$.
Hence, we may rewrite (\ref{eq:clustering}) as
\begin{align}
\label{eq:tau-clustering-simpler}
(\tau_j\mid \boldsymbol{\tau}_{\setminus j},\boldsymbol{Y},\boldsymbol{\Theta}) &\sim q_0 P_j(\tau_j) +
\sum\limits_{k=1}^K q_k \delta_{\eta_k}(\tau_j), 
\end{align}
where $K$ denotes the number of clusters, and $q_k$ is proportional to
\begin{align*}
&n^{(j)}_k \cdot f_{\mathcal{N}}(y_j-\mu_j;\mu_c/\sqrt{\eta_k},\sigma^2_c/\eta_k)
\end{align*}
with
$n^{(j)}_k =\left| \left\{j^{\prime} \ne j\::\: z_{j^{\prime}}=k \right\}
\right|.$
Rewriting (\ref{eq:tau-clustering-simpler}) using the
conditional cluster assignments gives
\begin{align}
\label{eq:clusteringonly}
(z_j\mid \boldsymbol{z}_{\setminus j},\boldsymbol{\eta},\boldsymbol{Y},\boldsymbol{\Theta}) &\sim q_0 \delta_{z_{new}}(z_j) +
\sum\limits_{k=1}^ K q_k \delta_k (z_j),
\end{align}
where $z_{new}=K+1$ unless $n^{(j)}_{z_j}=0$ in which case $z_{new}=z_j$.

The conditional in (\ref{eq:clusteringonly}) describes the assignment of
one node to a cluster given all the other cluster assignments and cluster
values.  We can also derive the conditional distribution of one cluster
value given all the other cluster values and all the cluster assignments.
Let $(k)\mathrel{\mathop:}= \{j:z_j=k\}$ and $n_k=|(k)|$.  The conditional
density (derived in the Appendix) is 
\begin{align}
\label{eq:clusterdraw}
f(\eta_k\mid \boldsymbol{\eta}_{\setminus k},\boldsymbol{Y},\boldsymbol{\Theta},\boldsymbol{z}) &= C(\alpha,\beta,\gamma)\cdot \eta_k^{\alpha - 1} \exp \left\{-\eta_k \beta - \sqrt{\eta_k} \gamma \right\} 
\end{align}
with
\begin{align*}
\alpha &= \frac{\nu+n_k}{2}, 
& \beta &= \frac{\nu + tr(\boldsymbol{\Theta}_{(k) (k) }\boldsymbol{Y}_{(k)} \boldsymbol{Y}^T_{(k)} )}{2}, 
& \gamma &= tr(\boldsymbol{\Theta}_{(k) \setminus (k)} \boldsymbol{X}_{\setminus (k)} \boldsymbol{Y}^T_{(k)}).
\end{align*}
The density being similar to (\ref{eq:full-conditional}), sampling can be
performed using the same rejection sampling scheme.  When the number of
clusters is small relative to $p$, cycling through the clusters and drawing
values for the whole cluster is much faster than cycling through all $p$
nodes and assigning each to a cluster (and drawing values when new clusters
are formed).

In the algorithm we may repeat some steps more frequently than others.  For
instance, if we are able to quickly identify clusters representing
significant deviations from normality, then we can perform the third step
more frequently than the computationally expensive reclustering step.
For initial values, we use the $tlasso$ of \cite{finegold} to estimate
$\boldsymbol{\tau}$, which means we have one cluster for each
observation.  

\subsection{Inference for the Clustering Parameter $\alpha$}

The Dirichlet Process parameter $\alpha$ plays a key role in
determining the number of clusters, and it is beneficial to add
another level of hierarchy and place a prior on $\alpha$.
Following \cite{escobar} who treat Dirichlet process mixture models,
we consider a $\Gamma(a,b)$ prior on $\alpha$.  In practice we choose
$a$ and $b$ to give low prior mean to $\alpha$, which leads to fewer
clusters and easier computation, but allows for more clusters as the
data requires.

Let $k$ denote the number of clusters.  We have from \cite{antoniak}
that, for $k=1,\dots,n$,
\begin{align}
P(k\mid \alpha,n) &= c_n(k)n! \alpha^k
\frac{\Gamma(\alpha)}{\Gamma(\alpha+n)}, 
\end{align}
where $c_n(k)=P(k\mid \alpha=1,n)$. Posterior inference is simplified by the fact that $\alpha$ is conditionally independent of the observed data given the cluster assignments, leading to
\begin{align}
P(\alpha\mid k,\pi,\boldsymbol{X}) &= P(\alpha\mid k) \propto P(\alpha)P(k\mid \alpha).
\end{align}
That is, inference is based on the prior for $\alpha$ and a single observation from $P(k\mid \alpha,n)$.

Returning to our setting with an $n$-sample, let the vector
$\boldsymbol{k}=(k_1,\dots,k_n)$ comprise the numbers of clusters in the $n$
observations.  Then $\alpha$ is conditionally independent of
$(\boldsymbol{\tau},\boldsymbol{Y},\boldsymbol{\Theta},\boldsymbol{\mu},G)$ given $\boldsymbol{k}$, and
\begin{align*}
P(\alpha\mid \boldsymbol{k},\boldsymbol{\tau},\boldsymbol{Y},\boldsymbol{\Theta},\boldsymbol{\mu},G) 
&\propto P(\alpha)P(\boldsymbol{k}\mid \alpha) \nonumber =P(\alpha)\prod\limits_{i=1}^n
P(k_i\mid \alpha). 
\end{align*}
Let $\beta(\alpha,\beta)$ be the Beta function. Generalizing the results of
\cite{escobar} to multiple observations $k_i$, we use the fact that
\begin{align}
\frac{\Gamma(\alpha)}{\Gamma(\alpha+p)} &= \frac{(\alpha+n) \beta(\alpha +1,p)}{\alpha \Gamma(p)}
\end{align}
to obtain that $P(\alpha\mid \boldsymbol{k})$ is proportional to
\begin{align}
&P(\alpha) \alpha^{\sum k_i -n } (\alpha+p)^n \int\limits_0^1
w_1^{\alpha} (1-w_1)^{p-1}dw_1 \times \cdots\times \int\limits_0^1 w_n^{\alpha} (1-w_n)^{p-1}dw_n.
\end{align}
Hence, we may view $P(\alpha\mid \boldsymbol{k})$ as a marginal
distribution of $P(\alpha,w_1,\dots,w_n\mid \boldsymbol{k})$ where
$0<w_i<1$ are random variables that are conditionally independent of
each other given $\alpha$.  Writing $\boldsymbol{w}=(w_1,\dots,w_n)$,
we consider the conditional distribution
\begin{align*}
P(\alpha\mid \boldsymbol{w},\boldsymbol{k}) &\propto \alpha^{a + \sum_{i=1}^n k_i-n-1} (\alpha+p)^n \exp
\bigg\{-\alpha\bigg(b-\sum_{i=1}^n \log w_i\bigg)\bigg\}. \nonumber  
\end{align*}
Expanding $(\alpha+p)^n$ gives a mixture of $n+1$ Gamma-distributions,
namely,
\begin{align}
\label{eq:condalpha}
(\alpha\mid \boldsymbol{w},\boldsymbol{k}) &\sim \sum_{j=0}^n \pi_j \Gamma \bigg(a+
  \sum_{i=1}^n k_i 
  -j,b - \sum_{i=1}^n \log w_i\bigg)  
\end{align}
with
\begin{align}
\pi_j &\propto {n \choose j}\, p^j\, \bigg(b - \sum_{i=1}^n \log w_i\bigg)^j\,
\Gamma\bigg(a + \sum_{i=1}^n k_i -j\bigg). 
\end{align}
The $w_i$ being conditionally independent given $\alpha$ and $\boldsymbol{k}$, it holds
that
\begin{align}
\label{eq:condw}
(w_i\mid \alpha,\boldsymbol{k},\boldsymbol{w}_{\setminus i}) &\sim \beta(\alpha+1,p).
\end{align}

Now suppose we observe $n$ independent random vectors
$\boldsymbol{Y}_1,\dots, \boldsymbol{Y}_n\in\mathbb{R}^p$ that follow
a $t^{\alpha}_{p,\nu}(\boldsymbol{\mu},\boldsymbol{\Psi})$
distribution.  Once again, let $\boldsymbol{Y}$ and $\boldsymbol{\tau}$
be the associated matrices of observations and divisors.  For small
$\alpha$, i.e., few expected clusters, we can create a sampler as
follows.  Let $k_i$ be the number of clusters for the $i^{th}$
observation.  The state space consists of the values for
$(G,\boldsymbol{\Theta},\boldsymbol{z},\boldsymbol{\eta})$ where
$\boldsymbol{z} = \{z_{ij}\}$ is now an $n \times p$ matrix and
$\boldsymbol{\eta}$ an array collecting $n$ vectors of length
$k_1,\dots,k_n$.  Following \cite{teh}, we propose the following Gibbs
sampler.

\begin{algorithm}[Dirichlet $t$]
  \label{alg:gibbsdirichlet}
  Starting with a decomposable graph $G_0$, and initial values
  $\boldsymbol{\mu}_0,$ $\boldsymbol{z}_0$, $\alpha_0$, and
  $\boldsymbol{\eta}_0$, repeat the following steps for
  $t=0,1,2,\dots$: \\ 
  (i) Jointly draw a new graph $G_{t+1}$ and a new matrix $\boldsymbol{\Theta}_{t+1}$ as in Algorithm \ref{alg:gibbsclassical}.\\
  (ii) For each observation $i=1,\dots,n$, cycle through the variables
  $j=1,\dots,p$ and draw $z_{ij}$ from the conditional given
  in (\ref{eq:clusteringonly}).  If $z_{ij}=z_{new}$ assign to this
  new cluster a value $\eta_{iz_{new}}$ by sampling from $P_j(\tau_j)$ in
  (\ref{eq:clustering}).  This results in a new matrix $\boldsymbol{z}_{t+1}$.\\
  (iii) For each observation $i=1,\dots,n$, cycle through the clusters
  $k=1,\dots,K_i$ and draw $\eta_{ik}$ using (\ref{eq:clusterdraw}).  This
  results in a new array $\boldsymbol{\eta}_{t+1}$.\\ 
  (iv) Assign $\boldsymbol{\mu}_{t+1}$ as in Algorithm \ref{alg:gibbsalternative}.\\
  (v) For each observation $i=1,\dots,n$, draw $w_i$ from the conditional given in (\ref{eq:condw}).\\
  (vi) Draw $\alpha$ from the conditional given in (\ref{eq:condalpha}).
\end{algorithm}

\section{Simulations}
\label{sec:simbayes}

\subsection{AR1 with $p$=25}
\label{sec:sim1}

To illustrate the behaviour of the different Bayesian methods, we
first present simulations for graphs with $25$ nodes, for which we run
the Markov chain Monte Carlo samplers for $10,000$ iterations per
possible edge, as suggested in \cite{carvalhostochastic}.  We choose a
graph for an autoregressive process of order one, that is, the nodes
form a chain and the corresponding precision matrix
$\boldsymbol{\Theta}$ is tri-diagonal.  We forego simulating random
draws of $\boldsymbol{\Theta}$ for a clear distinction between true
positives and negatives.  Instead, we set the non-zero off-diagonal
elements of $\boldsymbol{\Theta}$ to $-1$ and the diagonal elements to
$3$ (except the first and last, which are set to $2$).  Unless
otherwise noted, we fix the degrees of freedom $\nu =3$, the graph
prior parameter $d = 0.05$, recall (\ref{eq:graph-prior-d}), and the
hyperparameter $\delta =1$.  It is not desirable, however, to have
exactly the same priors in the normal and $t$-models.  With the most
extreme observations downweighted, the entries of the matrix
$\boldsymbol{S}_{\boldsymbol{\tau}\boldsymbol{Y}\boldsymbol{Y}}$ from
the $t$-model tend to be smaller in magnitude than those of the sample
covariance matrix $\boldsymbol{S}$, and, in our experience, the same
hyperparameter matrix $\boldsymbol{\Phi}$ then leads to larger graphs
for the $t$-case.  To get graphs of comparable size, we choose
$\boldsymbol{\Phi}=\boldsymbol{\mathcal{I}}_p/5$ for the normal model
and $\boldsymbol{\Phi}=\boldsymbol{\mathcal{I}}_p/10$ for $t$-models.

We first simulate $n$ independent normal observations from 
$\mathcal{N}_p(\boldsymbol{0},\boldsymbol{\Theta}^{-1})$.  In
order to illustrate the effects on inference of dependence structures, we
assume the mean to be known and run five different estimation methods:
\renewcommand{\theenumi}{\alph{enumi}}
\renewcommand{\labelenumi}{(\theenumi)}
\begin{enumerate}
\item The normal procedure (\S \ref{sec:armstrong}).
\item The normal procedure using the maximum likelihood estimate
  $\boldsymbol{S}_{\hat{\boldsymbol{\tau}}\boldsymbol{YY}}$ from the
  classical $t_{p,3}$-model as the sufficient statistic instead of
  $\boldsymbol{S}$.
\item The classical $t_{p,3}$ procedure
  (\S\ref{sec:bayesclassical}), drawing the matrix
  $\boldsymbol{\tau}$ 
  once every $10$ edge proposals.
\item The Dirichlet $t_{p,3}^{\alpha}$ procedure
  (\S\ref{sec:dirichlett}) with a $\Gamma(1,1)$ prior on $\alpha$.  We
  draw new cluster identifiers $\boldsymbol{z}$ for every $20$
  draws of $\boldsymbol{\eta}$, which in turn is drawn once every $10$
  edge proposals.
\item The alternative $t_{p,3}^*$ procedure (\S\ref{sec:altgibbs}),
  drawing the matrix $\boldsymbol{\tau}$ once every $10$ edge
  proposals.
\end{enumerate}

Assuming known means only favors normal procedures for which
estimation of the mean from heavy-tailed data is problematic in itself.
For each method, we perform $3$ million edge proposals.  We repeat the
process $50$ times, each time recording the posterior probability
$P(e_{jk}=1\mid \boldsymbol{Y})$, that is, the probability of edge
$\{j,k\}$ to be in the true edge set.  If $P(e_{jk}=1\mid \boldsymbol{Y})>
\epsilon$ for some threshold $\epsilon$, we consider it a ``positive''. We
let $\epsilon$ range from $0$ to $1$ and record the number of true and
false positives in all $50$ simulations.  We then compare the true positive
rate to the false positive rate at each threshold.  This entire process is
repeated for data generated from a
$t_{p,3}(\boldsymbol{0},\boldsymbol{\Theta}^{-1})$ and a
$t^*_{p,3}(\boldsymbol{0},\boldsymbol{\Theta}^{-1})$ distribution.

\begin{figure}
\centering
\includegraphics[width=6.9cm]{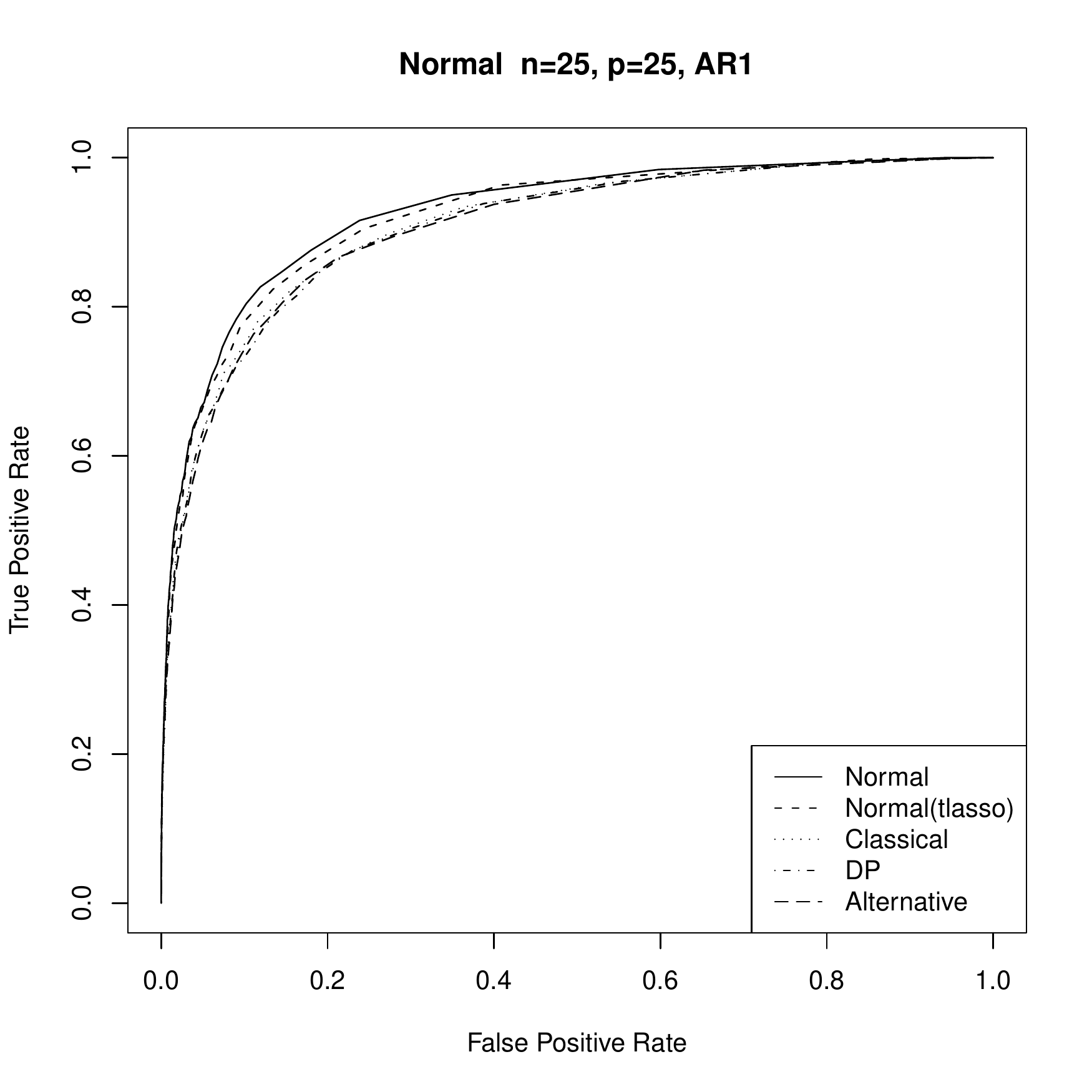}
\includegraphics[width=6.9cm]{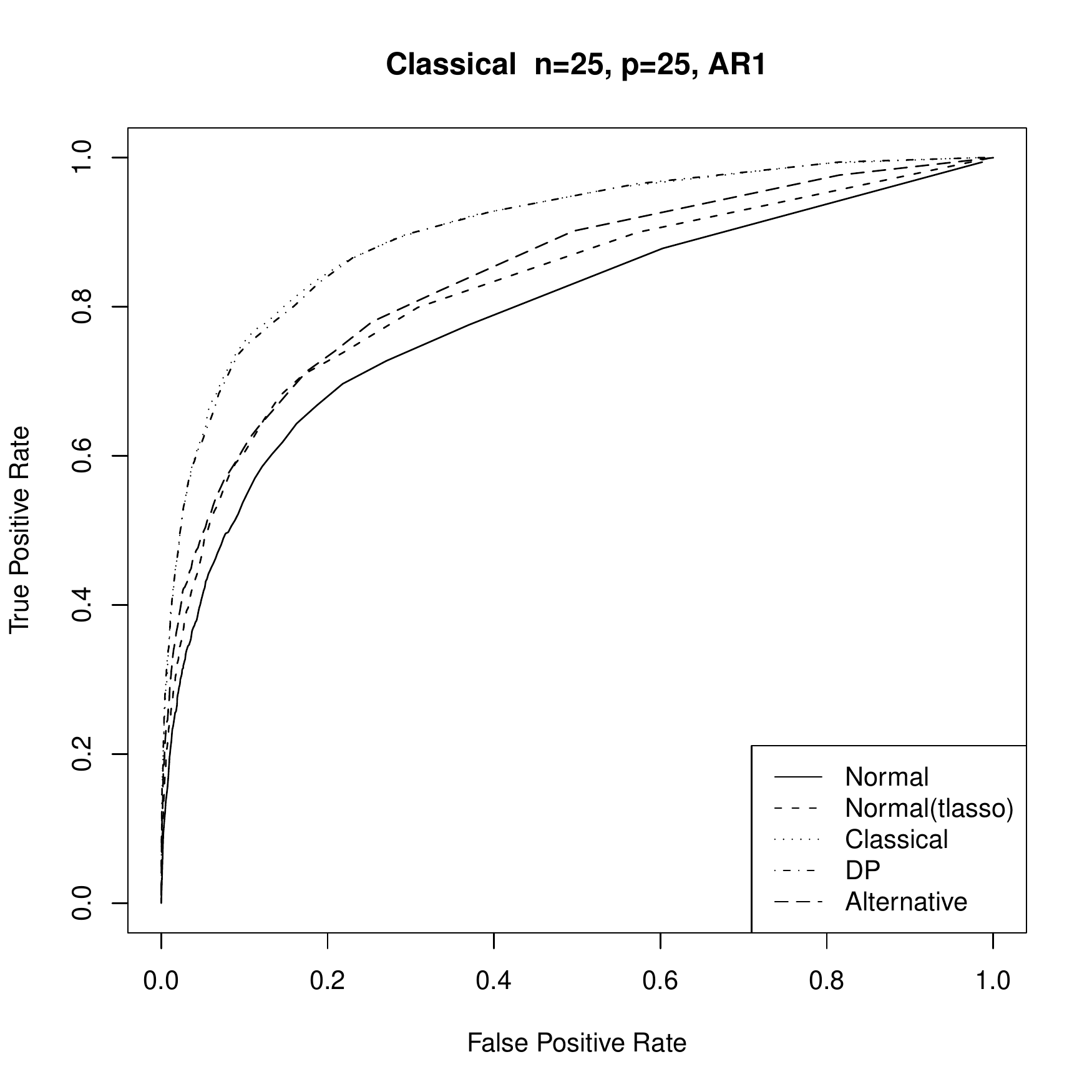}
\includegraphics[width=6.9cm]{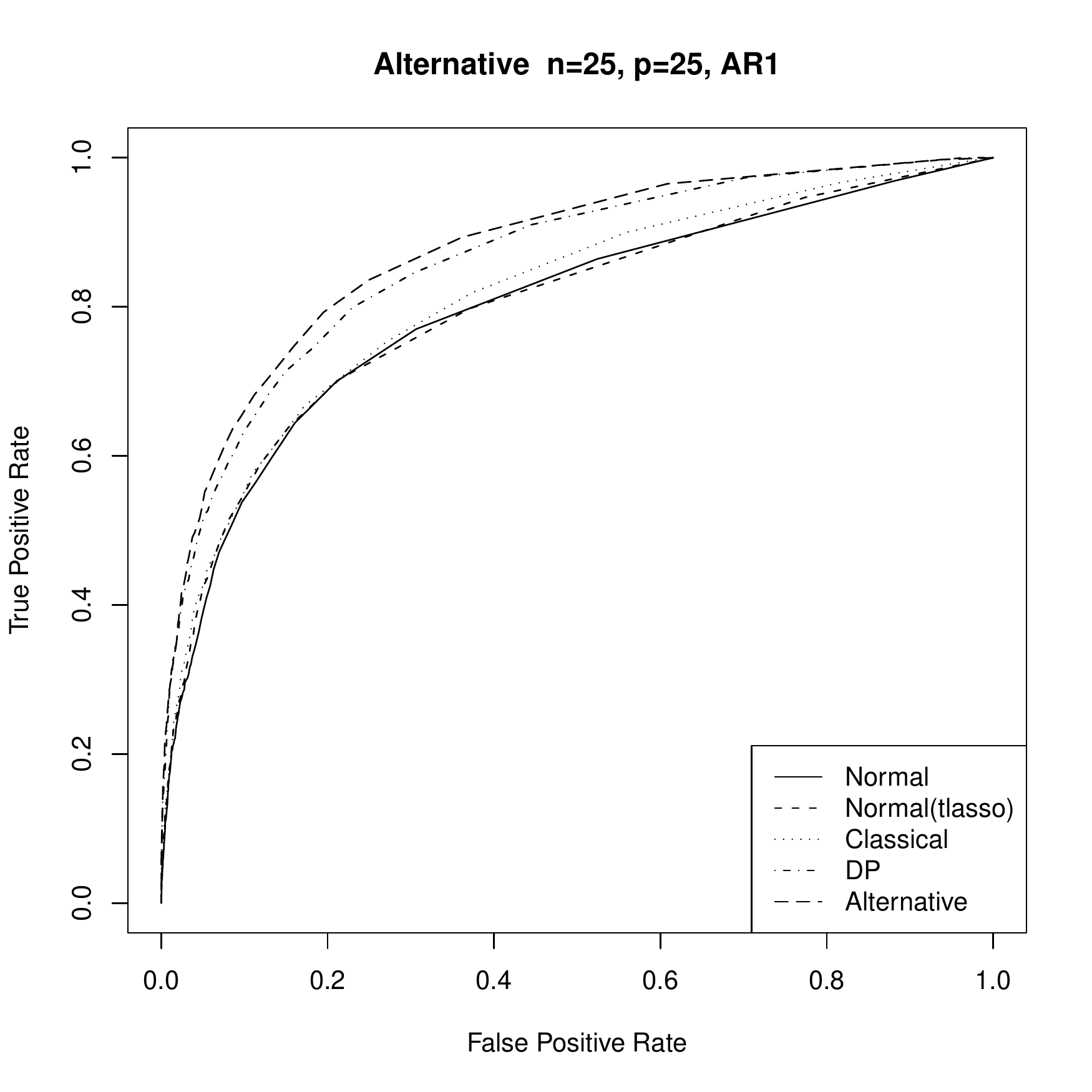}
\caption{ROC curves depicting the performances of the five methods for data generated from a $\mathcal{N}_{25}(\boldsymbol{0},\boldsymbol{\Theta}^{-1})$ distribution, a $t_{25,3}(\boldsymbol{0},\boldsymbol{\Theta}^{-1})$ distribution and a $t^*_{25,3}(\boldsymbol{0},\boldsymbol{\Theta}^{-1})$ distribution.}
\label{fig:bayesrocclassical} 
\end{figure}

The simulation results are summarized in Figure
\ref{fig:bayesrocclassical}, which shows that the more flexible models
indeed perform better when the data is generated from the more
complicated model.  With normal data, the $t$-models all perform
similarly well and the normal model outperforms them only slightly.
For classical $t$-data, the classical $t$-model performs significantly
better than the normal model.  The alternative model is clearly
inferior to the classical model.  The performance of the Dirichlet
$t$-model, on the other hand, is barely distinguishable from that of
the classical $t$-model.  With a $\Gamma(1,1)$ prior for $\alpha$, the
Dirichlet method finds an average of  $1.4$ clusters per
$\boldsymbol{\tau}$ vector, rendering it very similar to the classical
model.  The normal procedure with the robustified estimate
$\boldsymbol{S}_{\hat{\boldsymbol{\tau}}\boldsymbol{YY}}$ performs
better than the purely normal technique, but not as well as the
classical $t$-model.  In addition to the superior performance seen
here, the fully Bayesian approach has, of course, the added benefit of
a posterior distribution on $\boldsymbol{\tau}$.  This can be useful
to assess of each observation's consistency with normality.

For alternative $t$-data, the alternative $t$-method clearly outperforms
its normal and classical $t$-analogues, which perform equally poorly.  Only
the Dirichlet $t$-model, which finds an average of $7.7$ clusters per
$\boldsymbol{\tau}$ vector, performs comparably to the alternative model.
Its strong performance on both classical and alternative data suggests that
the Dirichlet method is indeed an effective compromise between the two
other $t$-distribution techniques.

For classical $t$-data, fitting the classical, the Dirichlet and the
alternative $t$-model took on average $2.3N$, $3.6N$ and $4.5N$,
respectively, where $N$ is the processing time for the normal model.
Based on use of `R' \citep{R}, the times are meant only to be rough
estimates of actual computational complexity.  Nevertheless, the
comparison suggests that the Dirichlet approach adaptively produces
statistically efficient estimates while using a run-time about halfway
between that of the classical and alternative procedures.  For
alternative $t$-data, the Dirichlet model faces the added complexity
of reclustering steps without much benefit from any clustering.
Indeed, the average run time was $5N$ for the Dirichlet model compared
to $3.8N$ for the alternative model.  This said, we would not expect
any real application requiring as large a number of clusters as
simulated alternative $t$-data.

\subsection{Random Graphs with $p$=100}

For a more challenging scenario, we consider $p=100$ nodes and create
the graph by forming 20 random cliques of size 2 to 5.  For each
clique, we pick nodes at random and form edges between all nodes in
the clique.  We draw the mean vector $\boldsymbol{\mu}$ as a
$p$-vector of independent standard normals.  We set the non-zero
entries in $\Theta$ as before (but multiply the diagonal entries by a constant
to ensure a minimum eigenvalue of at least 0.6).  We then simulate $n=100$
independent observations from a
$\mathcal{N}_p(\boldsymbol{0},\boldsymbol{\Theta}^{-1})$ distribution
to create  latent data $X$.

Next, we contaminate the data via an $n \times p$ matrix
$\boldsymbol{\tau}$ that holds divisors for $X$, with the goal of
creating contamination in many parts of many observations so that
detecting outliers manually would be difficult.  For each one of a
total of 10 contaminations, we draw a Poisson number of rows and a
Poisson number of columns (mean 10 for both).  We then select
uniformly at random a submatrix of $\boldsymbol{\tau}$ that has this
given size and assign a single random value (uniform[0.01,0.2]) to the
entries in the submatrix.  The remaining entries of
$\boldsymbol{\tau}$ are set to 1.  The observations $Y$ are created by
setting $Y_{ij} = \mu_j+X_{ij}/\tau_{ij}$.  This results on average in
contamination in slightly less than one in ten elements of the latent
data matrix.

\begin{figure}
\centering
\includegraphics[width=13cm]{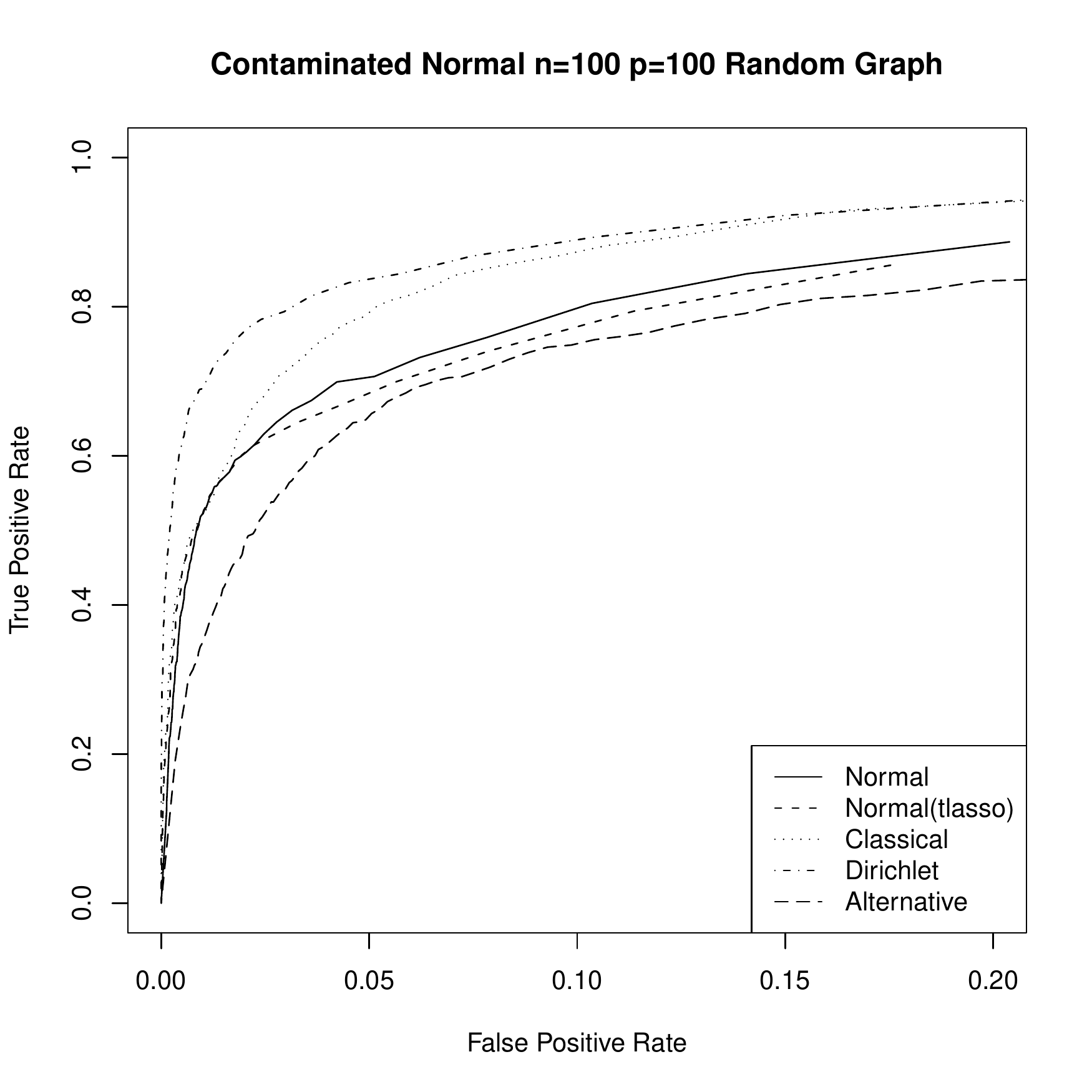}
\caption{ROC curves depicting the performances of the five methods for data generated from a contaminated $\mathcal{N}_{100}(\boldsymbol{0},\boldsymbol{\Theta}^{-1})$ distribution that is Markov to a random graph.}
\label{fig:bayesroccont} 
\end{figure}

We run the five algorithms from Section~\ref{sec:sim1} under the
settings described there, but with
$\boldsymbol{\Phi}=\boldsymbol{\mathcal{I}}_p/5$ for the normal model
and $\boldsymbol{\Phi}=\boldsymbol{\mathcal{I}}_p/20$ for $t$-models,
to get graphs of comparable size.  We run the samplers to obtain
2 million edge draws and for the $t$ models we sample
$\boldsymbol{\tau}$ every 30 edge draws.  The results for 25 repeats of the entire
process are shown in Figure~\ref{fig:bayesroccont}, which makes a
clear case for the Dirichlet $t$-model.

%%%%%%%%%%%%

\begin{figure}
\centering
\includegraphics[width=5cm]{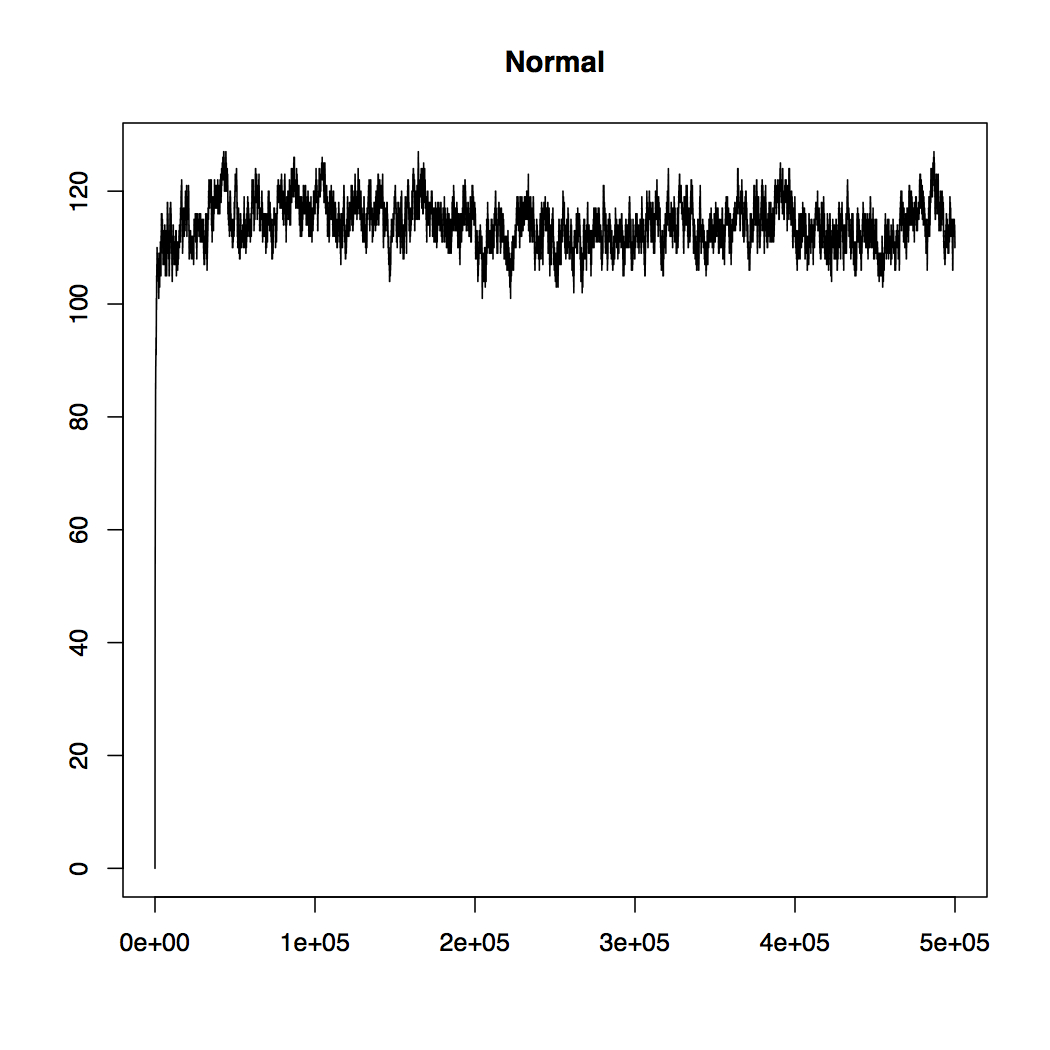}
\includegraphics[width= 5cm]{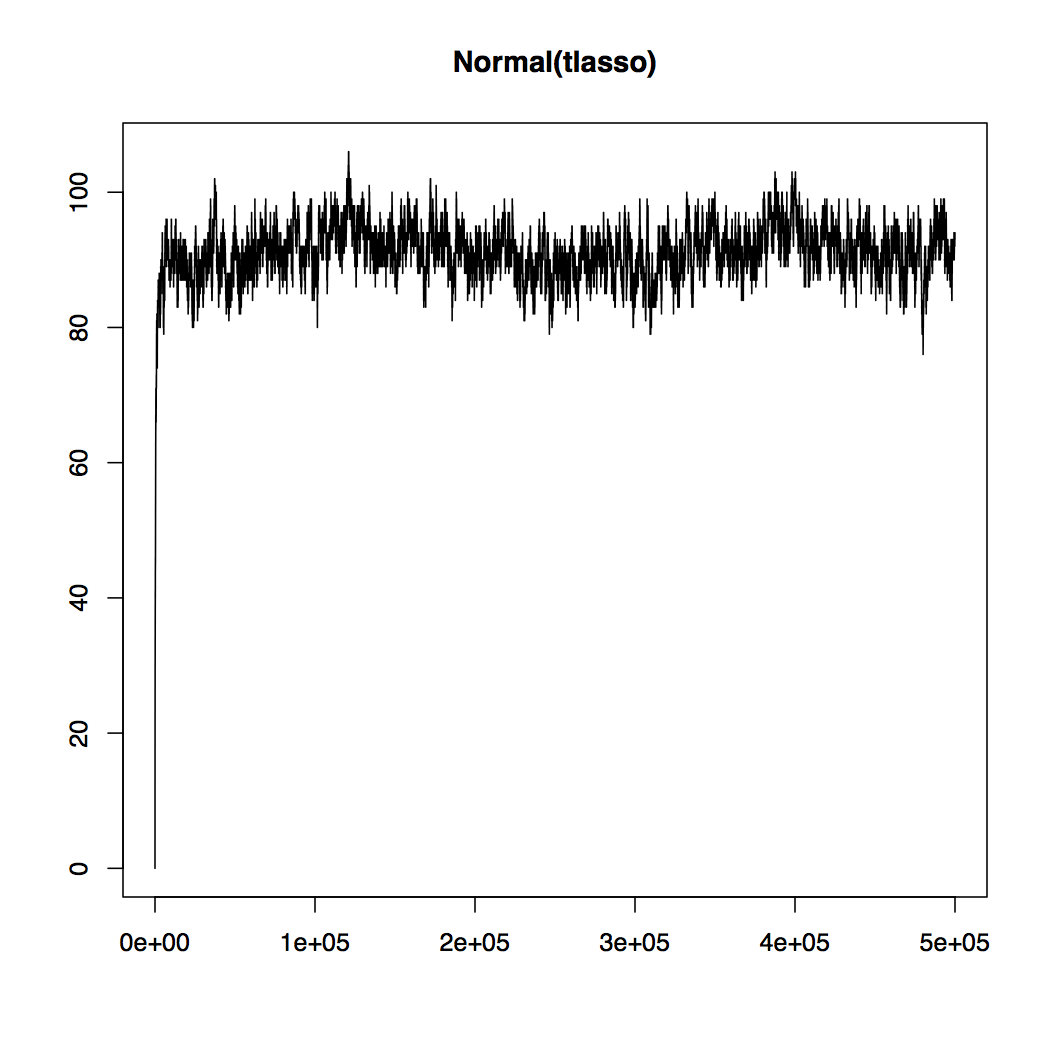}
\includegraphics[width= 5cm]{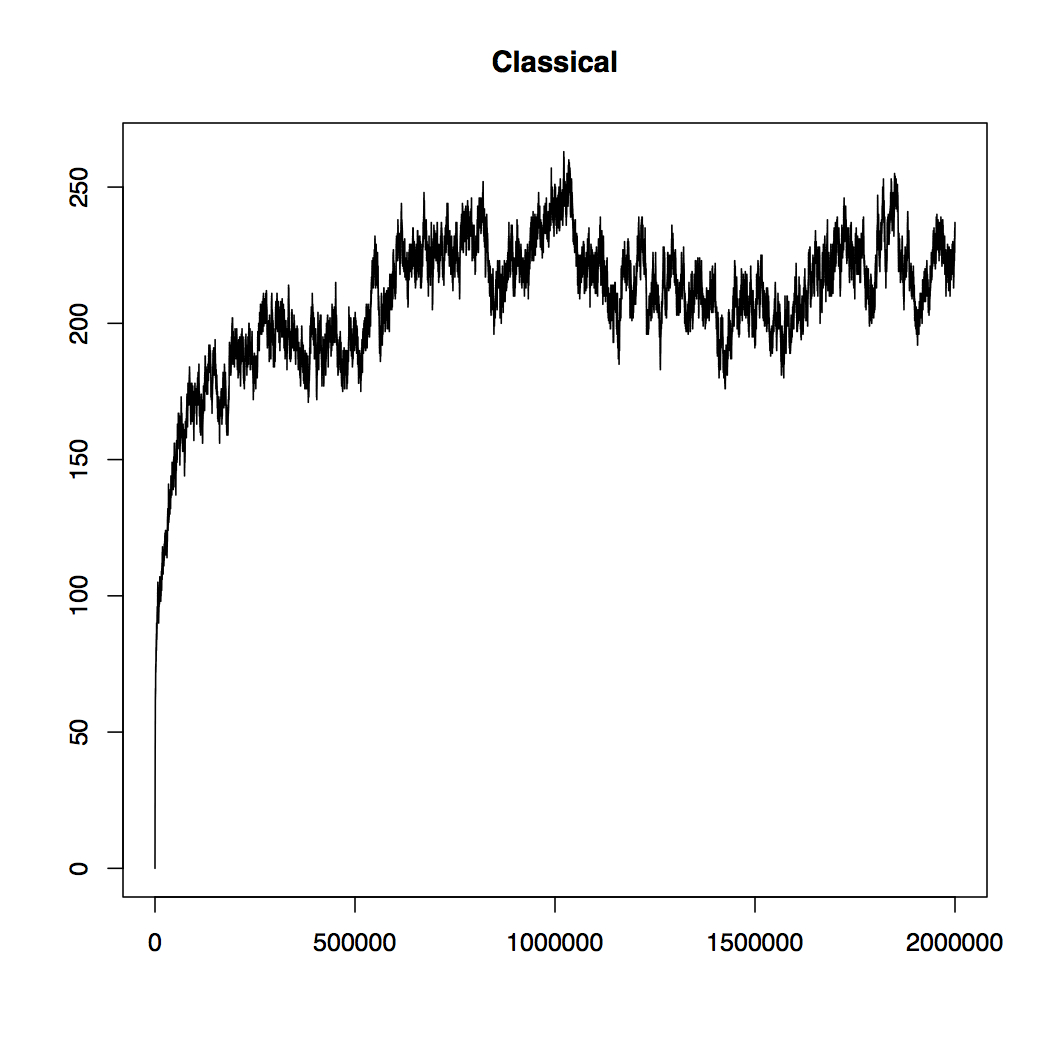}
\includegraphics[width= 5cm]{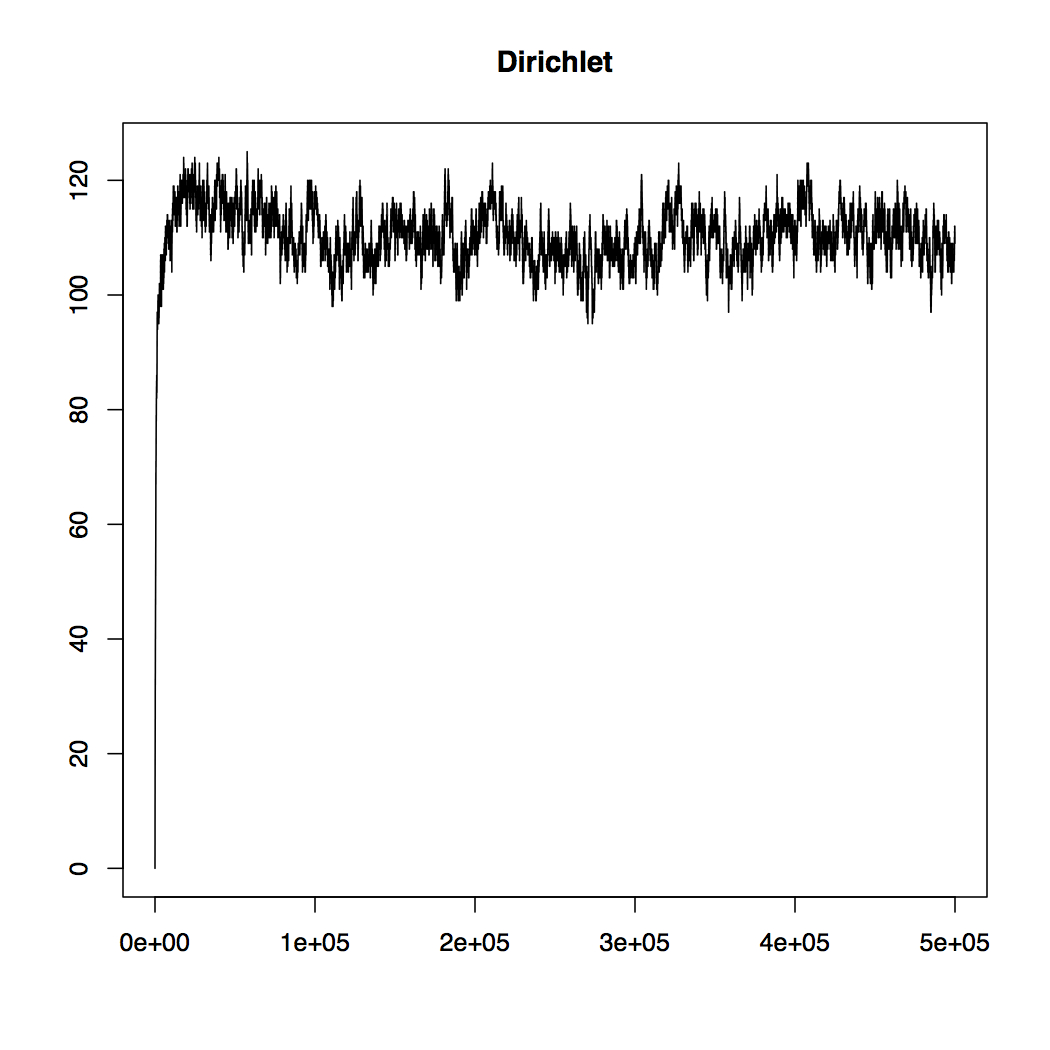}
\includegraphics[width= 5cm]{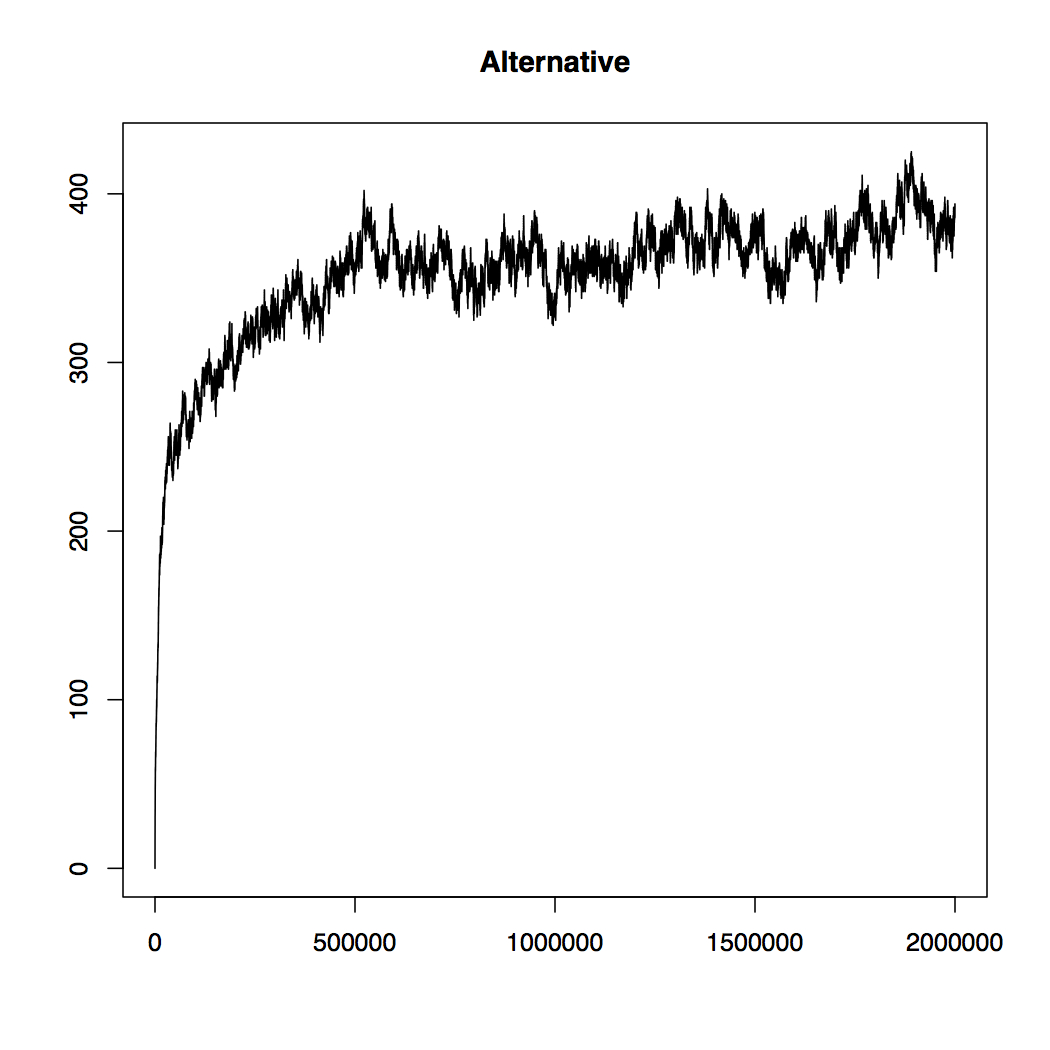}
\caption{Edges in estimated graph plotted against iterations of the
  Gibbs sampler for models used in the contaminated normal
  simulations.} 
\label{fig:edges} 
\end{figure}

The 2 million draws do not seem enough for
``convergence'' of all Markov chains.  Figure \ref{fig:edges}, which
shows the number of edges in the estimated graph over the iterations
of the samplers, suggests that the alternative model may require much
longer runs.  As another plus, the
Dirichlet model seems to require far fewer iterations.

\section{Gene Expression Data}
\label{sec:bayesgasch}

\begin{figure}[t]
\centering
\vspace{-.5cm}
\hspace{-.7cm}
\includegraphics[width=10cm]{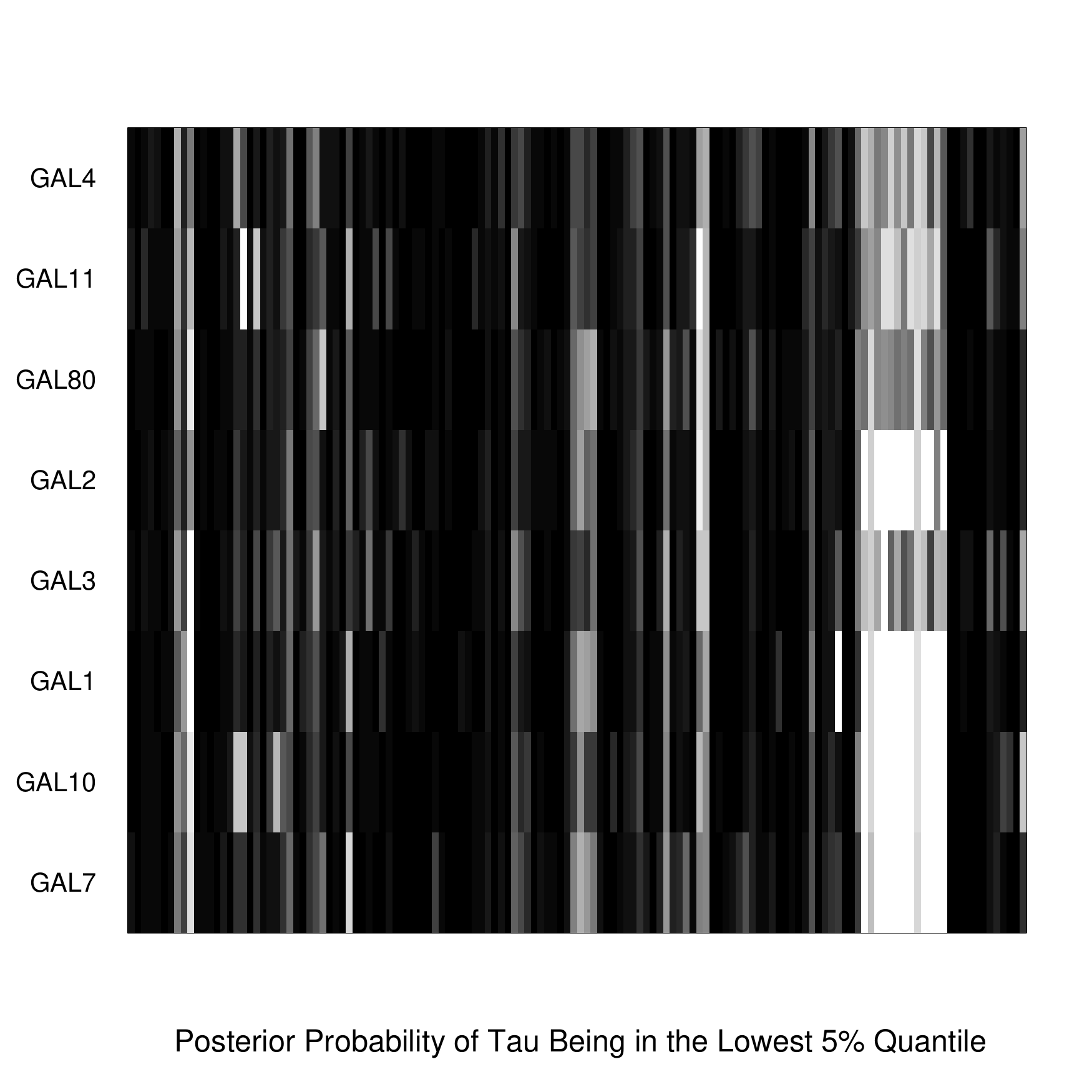}
\caption{Lighter colors represent higher posterior probability that
  $\tau_{ij}$ is in the bottom $5\%$ quantile of the prior
  distribution on $\tau$.  The light blocks on the bottom right
  correspond to the ``outliers'' in the yeast data from Section
  \ref{sec:bayesgasch}.} 
\label{fig:heatmappost}
\end{figure}

\cite{gasch} present data from microarray experiments with yeast
strands.  As in the frequentist work in \cite{finegold}, we focus on
$8$ genes involved in galactose utilization; 136 experiments have data
for all the 8 genes.  In $11$ experiments, 4 of the genes have
abnormally large negative expression values.
Figure~\ref{fig:heatmappost}, which is obtained from $3$ million
iterations of our sampler, shows that the Dirichlet
$t^{\alpha=1}_{3,8}$ model leads to downweighting of the abnormal
values but not entire observations, as desired.

\begin{figure}
\centering
\includegraphics[width=5cm]{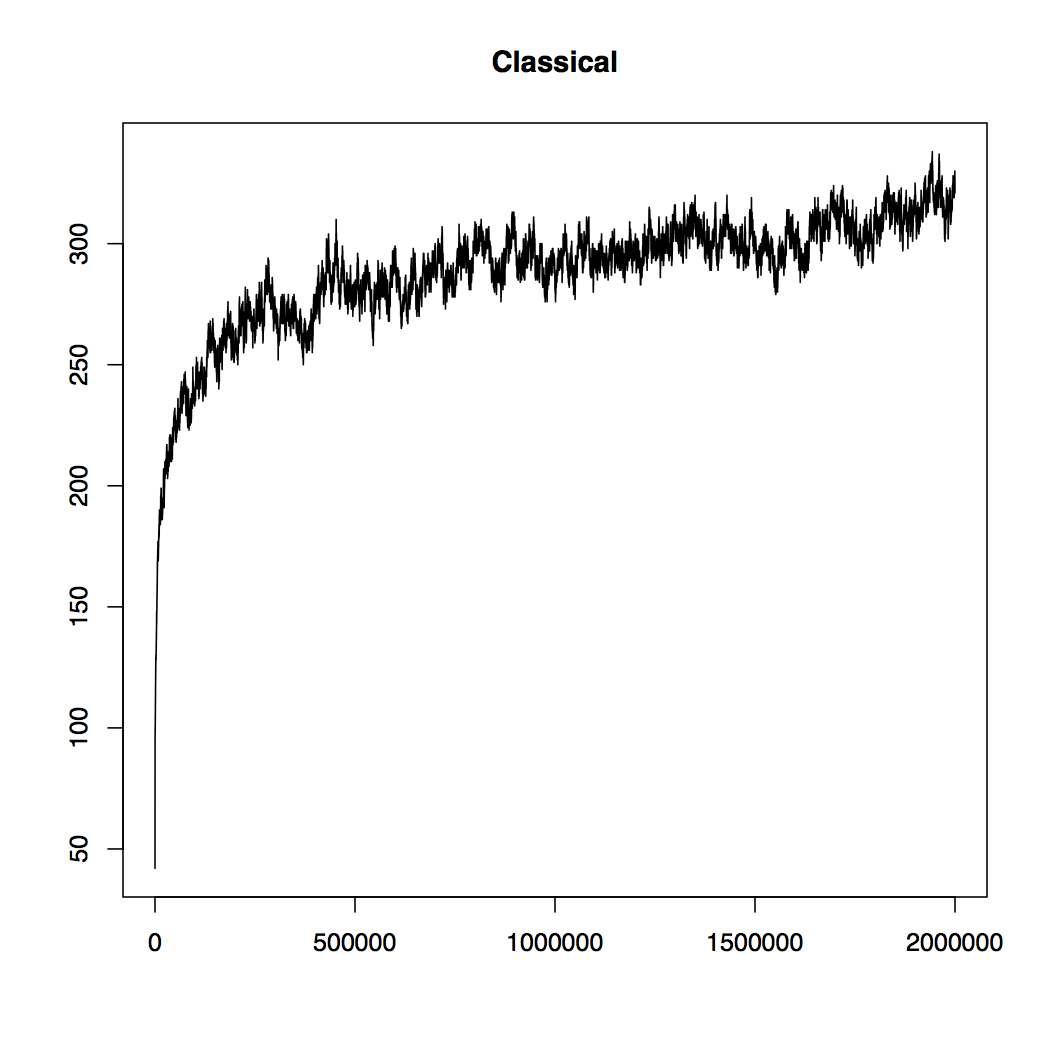}
\includegraphics[width=5cm]{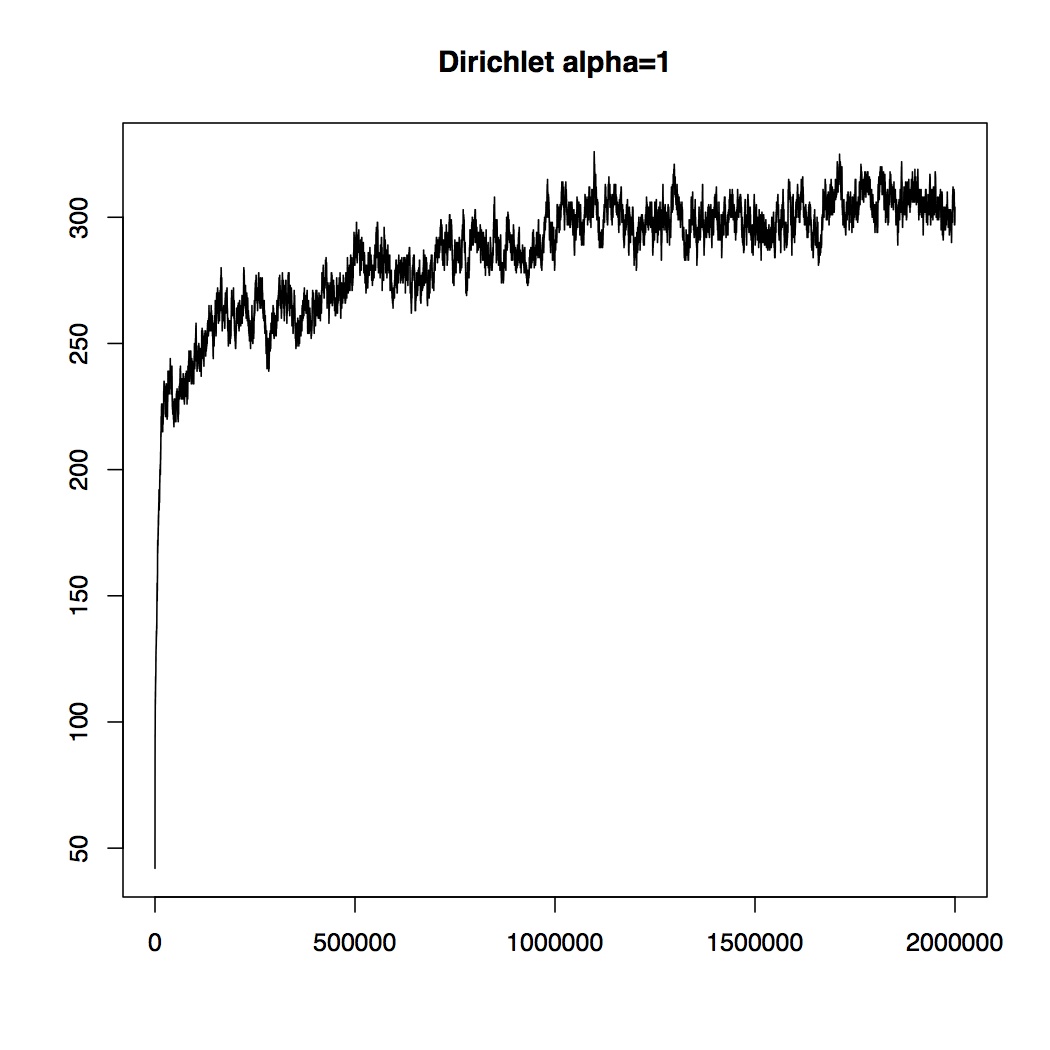}\\
\includegraphics[width=5cm]{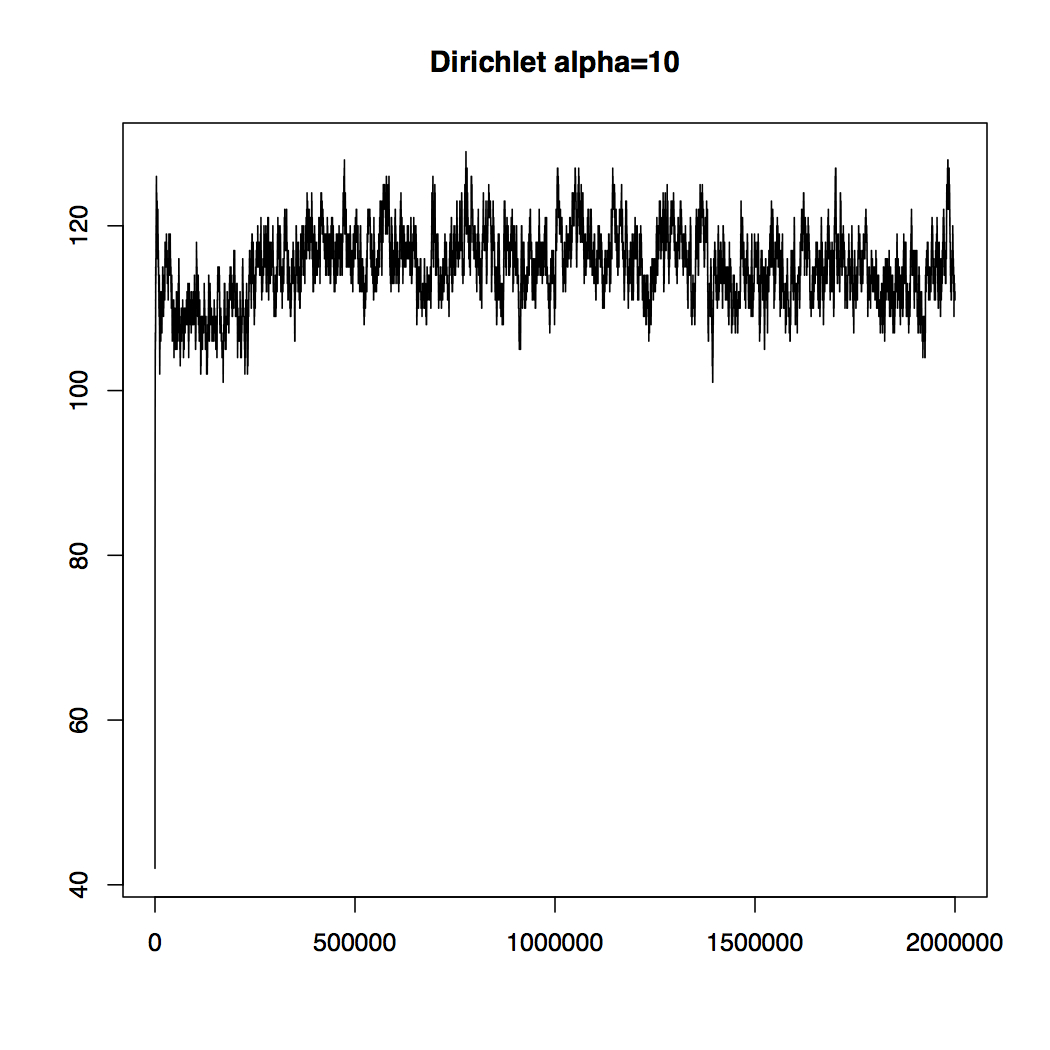}
\includegraphics[width=5cm]{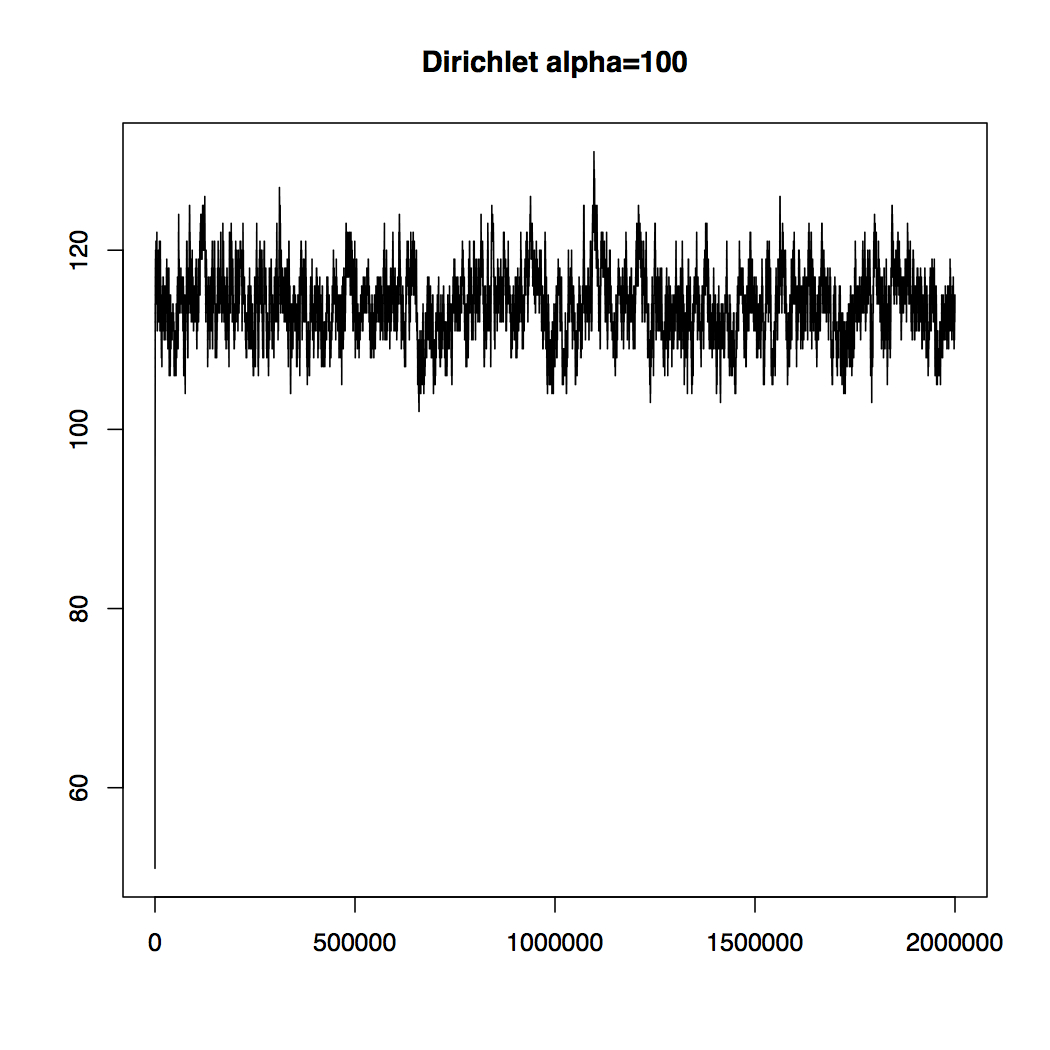}
\caption{Number of edges in estimated graph plotted against iterations
  of the Gibbs sampler for models used in analysis of gene expression
  data. } 
\label{fig:gasch.edges} 
\end{figure}

To demonstrate automatic detection of outliers in a much larger
dataset, we add the data for 92 genes, selected at random from those
without missing data.
We fit the models $t^{\alpha}_{3,100}$ for fixed
$\alpha=\{1,10,100\}$.  We run the samplers for $2$ million edge
draws
and 
follow a proposal of \cite{donnet} to select
candidate edges with probability proportional to the sample
correlation of the corresponding variables.  Figure
\ref{fig:gasch.edges} provides convergence diagnostics.

For $\alpha=100$ the propensity to cluster is very weak with an
average of 67 different clusters in each row of $\boldsymbol{\tau}$ in
the final iteration.  Nonetheless, the average $\boldsymbol{\tau}$
value for the 4 relevant genes in the 11 relevant experiments was
0.046 compared with an average value of 0.84 for the rest of the
matrix.  That is, as we let the Dirichlet $t$ model approach the
alternative model, we achieve downweights of the suspected outliers
even without the benefits of clustering.

With $\alpha=10$ the propensity to cluster is a bit stronger; the rows
of $\boldsymbol{\tau}$ have an average of 20 different clusters in the
final iteration.  We achieve relative downweighting similar to the
$\alpha=100$ model (0.067 to 0.47).  Despite 20 clusters per row, at
least two of the outliers always share a cluster and in five of the
experiments, at least three share a cluster.  

Letting $\alpha=1$, we achieve much greater clustering (2.8 per row).
The suspected outliers tend to cluster together, but there is little
practical difference between this model and the classical $t$.  The
classical $t$  downweights most of the observations
significantly and takes much longer to reach the appearance of
equilibrium in the estimated edges; see Figure~\ref{fig:gasch.edges}.

\section{Discussion}
\label{sec:discussion}

We have extended Bayesian approaches to graphical Gaussian
modeling using three variations of multivariate $t$-distributions.
While these extensions all come at increased computational expense,
they can have substantial statistical benefit.  In particular, one
obtains a posterior distribution for latent weights that measures
uncertainty about potential outliers.

Our extensions to $t$-distributions are based on one particular
Gaussian model, but many other variations have been treated in the
literature.  Some authors place a prior on the degrees of freedom
parameter $\delta$ for the Hyper Inverse Wishart (HIW) prior
distribution.  In addition, one can introduce a prior on the edge
density parameter $d$ from (\ref{eq:graph-prior-d}).  An alternative
approach is to place a uniform prior on the size of the graph, and
then a uniform prior on all graphs of the same size.  We have chosen
the scale matrix of the HIW prior to be $\boldsymbol{\Phi}=c
\boldsymbol{\mathcal{I}}_p$ and treated fixed choices of $c$ in our
simulations.  One could instead include a prior on $c$, or set
$\boldsymbol{\Phi}=c\boldsymbol{A}$, where $\boldsymbol{A}$ is the
sample covariance matrix, or an equicorrelated matrix with diagonal
elements equal to $1$ and off-diagonal elements equal to a common
value $\rho$.  \cite{armstrong} consider all of these variations,
including placing a prior on $\rho$.  Finally, as mentioned in the
introduction, there now exist more flexible versions of the HIW
distribution as well as techniques for approximate computations for
non-decomposable graphs.  Incorporating the former generalization in
the $t$-distribution context would be straightforward; addressing
non-decomposable graphs, however, constitutes an interesting area for
further research.

As noted previously by many authors, for large $p$ even the most
likely graphs may have very small posterior probability, making it
difficult and not necessarily very informative to identify the graph
with highest posterior probability.  In practice, the focus may thus
often be on more modest goals, such as the posterior distribution of
subgraphs on some subset of vertices, or even more simply, the
marginal posterior probability that each edge is in the edge set $E$
that we considered in the simulation study.

In the data from Section \ref{sec:bayesgasch}, a group of genes had
extreme expression values in several observations.  While the
Dirichlet $t$-model did a good job identifying these clusters, it
could potentially be worthwhile to share statistical strength by
explicitly modeling the clustering of latent weights as similar across
observations.  This could be done by treating the $p$-vectors
$\tau_1,\dots,\tau_n$ as draws from a Dirichlet process mixture model.
Combined with the Dirichlet $t$-model, this would give a `doubly Dirichlet'
$\boldsymbol{\tau}$ matrix.  That is, we
will have $k \le p$ distinct elements within each row (observation)
and $l \le n$ distinct rows.  Inference in this model would be more
involved than in the ordinary Dirichlet $t$-model we discussed, but
the full conditionals necessary to devise a Gibbs sampler would be of
similar type.

\section*{Acknowledgements}

This work was supported by the U.S.~National
Science Foundation under Grant No.~DMS-0746265.  Mathias Drton was
also supported by an Alfred P.~Sloan Fellowship.

\bibliographystyle{biom}
\bibliography{MFref}

\clearpage
\appendix

\section{Appendix}
\subsection{Alternative Procedures for Drawing the Mean Vector
  $\boldsymbol{\mu}$} 
\label{app:mutheor}

The parameters of the normal distribution in \eqref{eq:mupost} involve
the inverse of the sum of matrices $\sum_{i=1}^n \tau_i
\boldsymbol{\Theta} + \boldsymbol{\Theta_{\mu}}$, where
$\boldsymbol{\Theta_{\mu}}= \boldsymbol{\mathcal{I}}_p/\sigma_{\mu}$
is a multiple of the identity matrix.  With rare exceptions, the
hyperparameter $\sigma_{\mu}$, a variance, is chosen large so as to
make the prior distribution on $\boldsymbol{\mu}$ flat.  In those cases,
$\boldsymbol{\Theta_{\mu}}$ is small compared to the typical values of
$\sum_{i=1}^n \tau_i \boldsymbol{\Theta}$.  Hence, little is
lost by ignoring the term $\boldsymbol{\Theta_{\mu}}$ in the matrix
inversion, which leads to the distributional approximation
\begin{align}
\label{eq:mupost-approx}
(\boldsymbol{\mu}\mid
  \boldsymbol{Y},G,\boldsymbol{\tau},\boldsymbol{\Theta}) &\approx
  \mathcal{N}_p \left( 
  \frac{1}{\textstyle \sum_{i=1}^n
  \tau_i}\cdot {\sum_{i=1}^n \tau_i
  \boldsymbol{Y}_i},\frac{1}{\textstyle 
  \sum_{i=1}^n\tau_i}\cdot \boldsymbol{\Theta}^{-1} \right). 
\end{align}

This approximation still requires the completion step we have avoided
to obtain $\boldsymbol{\Theta}^{-1}=\boldsymbol{\Psi}$.  While this is
not prohibitively expensive, we find that simply setting
$\boldsymbol{\mu}$ equal to the mean of the distribution in 
(\ref{eq:mupost-approx}) works well enough in practice -- we call this
``Robust Centering''. To test this, we simulated classical $t$ data
from a chain graph with 25 nodes as described in Section
\ref{sec:simbayes}.  We ran four versions of the the Dirichlet $t$
algorithm (Algorithm 4) starting with the same seed: using naive
centering (subtract of the sample mean for each variable and set
$\boldsymbol{\mu}=\boldsymbol{0}$); Robust Centering; sampling from
the approximate conditional in equation (\ref{eq:mupost-approx}); and
sampling from the exact conditional in \eqref{eq:mupost}.  With
$\sigma_{\mu}$ set to 100,000 we find virtually no difference between
the estimated values of $\boldsymbol{\mu}$ from the last three
procedures, but significant difference between those three and the
first.  The first procedure does a worse job of estimating
$\boldsymbol{\mu}$ and, as a result, $\boldsymbol{\Theta}$.  We
conclude that Robust Centering is better than naive centering, but
that virtually nothing is lost by failing to sample from the
approximate or full conditionals.

\subsection{Full Conditional for Latent Divisors in the Dirichlet $t$-Model}
\label{app:conddir}

For notational convenience, let $z_j=0$ if $\tau_j$ belongs to a new
cluster and consider the case $j=p$.  Let $K$ be the number of
distinct clusters containing elements other than $\tau_p$. We may then
write
\begin{align*}
  P(\tau_p \le t \mid \boldsymbol{\tau}_{\setminus p},\boldsymbol{Y},\boldsymbol{\Theta}) &=
  \sum\limits_{k=0}^{K} P(\tau_p \le t\mid \boldsymbol{\tau}_{\setminus p},\boldsymbol{Y},z_p=k)
  P(z_p=k\mid \boldsymbol{\tau}_{\setminus p}, \boldsymbol{Y}, \boldsymbol{\Theta}).
\end{align*}
The conditional density of $\tau_p$ given $(\boldsymbol{\tau}_{\setminus p},
\boldsymbol{Y},\boldsymbol{\Theta},z=k)$ is trivially the point mass at $\eta_k$, the value
assumed by all elements of $\boldsymbol{\tau}$ that belong to the $k^{th}$ cluster.
The conditional density of $\tau_p$ given $(\boldsymbol{\tau}_{\setminus p},
\boldsymbol{Y},\boldsymbol{\Theta},z=0)$ is
\begin{align}
  f(\tau_p\mid \boldsymbol{\tau}_{\setminus p},\boldsymbol{Y},\boldsymbol{\Theta},z=0) &\propto f(\tau_p\mid \boldsymbol{\tau}_{\setminus p},\boldsymbol{\Theta},z=0)f(\boldsymbol{Y}\mid \boldsymbol{\tau},\boldsymbol{\Theta},z=0) \nonumber \\
  &=f(\tau_p\mid z=0)f(\boldsymbol{Y}\mid \boldsymbol{\tau},\boldsymbol{\Theta}) \nonumber \\
  \label{eq:full-cond-tau-z0}
  &\propto f_{\Gamma}(\tau_p;
  \nu/2,\nu/2)f_{\mathcal{N}}(Y_p;\mu_c/\sqrt{\tau_p},\sigma^2_c/\tau_p),
\end{align}
where $f_{\Gamma}(\tau_p; \nu/2,\nu/2)$ is the density of a
$\Gamma(\nu/2,\nu/2)$ distribution evaluated at $\tau_p$.  The
distribution specified by (\ref{eq:full-cond-tau-z0}) is the Gibbs
sampling distribution from the alternative model in
(\ref{eq:full-conditional}).  Now,
\begin{align*}
  P(z_p=0\mid \boldsymbol{\tau}_{\setminus p},\boldsymbol{Y},\boldsymbol{\Theta}) &\propto P(z_p=0,Y_p\mid
  \boldsymbol{Y}_{\setminus p},\boldsymbol{\tau}_{\setminus p},\boldsymbol{\Theta}) \nonumber \\
  &=P(z_p=0\mid \boldsymbol{\tau}_{\setminus p}, \boldsymbol{Y}_{\setminus p},\boldsymbol{\Theta}) P(Y_p\mid
  \boldsymbol{X}_{\setminus p},\boldsymbol{\Theta},z_p=0) \nonumber \\
  &\propto \alpha P(Y_p\mid \boldsymbol{X}_{\setminus p},\boldsymbol{\Theta},z_p=0).
\end{align*}
If $z_j=0$ then $X_j$ and $\tau_j$ are conditionally independent of
$(\boldsymbol{Y}_{\setminus j},\boldsymbol{\tau}_{\setminus j})$ and each other given
$X_{\setminus j}$.  Let $\mu_c$ and $\sigma_c$ be the conditional mean
and standard deviation of $(X_j\mid \boldsymbol{X}_{\setminus j})$.  Therefore, 
given $\boldsymbol{X}_{\setminus j}$, the random variable $Y_j/\sigma_c$ has the same
distribution as 
\begin{align*}
   \frac{\mu_c/\sigma_c+Z}{\sqrt{\tau_j}}
\end{align*}
for $Z \sim \mathcal{N}(0,1)$.  We recognize the distribution as a
noncentral $t$-distribution with degrees of freedom $\nu$ and
noncentrality parameter $\mu_c/\sigma_c$.

Similarly, for $k>0$,
 \begin{align*}
   P(z_p=k\mid \boldsymbol{\tau}_{\setminus p},\boldsymbol{Y},\boldsymbol{\Theta}) &\propto P(Y_p\mid
   \boldsymbol{\tau}_{\setminus p},\boldsymbol{Y}_{\setminus p},\boldsymbol{\Theta},z_p=k).
 \end{align*}
 Now, if $z_p=k$, then $Y_p=X_c/\sqrt{\tau_k}$ where $X_c \sim
 \mathcal{N}(\mu_c,\sigma_c)$.  Therefore, the conditional
 distribution of $Y_p$ given $\boldsymbol{\Theta},\boldsymbol{Y}_{\setminus p},
 \boldsymbol{\tau}_{\setminus_p},$ and $z_p=k$ is
\begin{align*}
f_{Y}(y) &= f_{X_c}(\sqrt{\tau_k}y)\sqrt{\tau_k}.
\end{align*}
Combining all the above elements gives the result stated in
(\ref{eq:clusteringonly}).

\subsection{Full Conditional for Cluster Values  in the Dirichlet $t$-Model
\label{app:condgivencluster}}

Define $(k)= \{j:z_j=k\}$ and
$\setminus(k)=\{1,\dots,p\}\setminus(k)$.  First, note that the pair
$(\boldsymbol{Y}_{(k)},\eta_{k})$ is conditionally independent of $\boldsymbol{Y}_{\setminus
  (k)}$ given $\boldsymbol{X}_{\setminus (k)}$.  Hence,
\begin{align*}
&f(\eta_k\mid \eta_{\setminus k},\boldsymbol{Y},\boldsymbol{\Theta},z)  = f(\eta_k\mid \boldsymbol{\boldsymbol{Y}}_{(k)},\boldsymbol{X}_{\setminus (k)},z) 
  \propto f(\eta_k,\boldsymbol{Y}_{(k)}\mid \boldsymbol{X}_{\setminus (k)},z).
\end{align*}
The last density is equal to 
\begin{align}
  \label{eq:app-gamma-normal}
  f(\eta_k\mid \boldsymbol{X}_{\setminus (k)})f\left(\boldsymbol{Y}_{(k)}\mid \boldsymbol{X}_{\setminus
  (k)},\eta_k,z \right) 
  &= f_{\Gamma}(\eta_k;\nu/2,\nu/2)\cdot f_{\mathcal{N}}\left(\boldsymbol{Y}_{(k)};\frac{\mu_{c}}{\sqrt{\eta_k}},\frac{\boldsymbol{\Sigma}_c}{\eta_k}\right) 
\end{align}
where $\boldsymbol{\mu}_c$ and $\boldsymbol{\Sigma}_c$ are the conditional
mean vector and covariance matrix of $\boldsymbol{X}_{(k)}$ given
$\boldsymbol{X}_{\setminus (k)}$.  By the well-known formulas for inverse
of a partitioned matrix, $\boldsymbol{\Theta}_{(k)(k)}$ is the inverse of
$\boldsymbol{\Sigma}_c$, and $\boldsymbol{\mu}_c$ is equal to
$\boldsymbol{\Theta}_{(k)(k)}\boldsymbol{\Theta}_{(k)\setminus(k)}\boldsymbol{X}_{\setminus(k)}$.The
product in (\ref{eq:app-gamma-normal}) is thus proportional to
\begin{align*}
  & \eta_k^{\nu/2 -1} \exp \{-\eta_k \nu/2\} |\eta_k
  \boldsymbol{\Theta}_{(k)(k)}|^{1/2} \exp \left\{-\frac{1}{2}
  \left(\boldsymbol{Y}_{(k)}-\frac{\boldsymbol{\mu}_c}{\sqrt{\eta_k}}\right)^T\eta_k
  \boldsymbol{\Theta}_{(k)(k)}\left(\boldsymbol{Y}_{(k)}-\frac{\boldsymbol{\mu}_c}{\sqrt{\eta_k}}\right)
  \right\} \nonumber \\ 
  &=\eta_k^{(\nu+|(k)|)/2 -1} \exp \left\{-\eta_k \left[\nu/2 +
  tr(\boldsymbol{\Theta}_{(k)(k)}\boldsymbol{Y}_{(k)}\boldsymbol{Y}^T_{(k)})/2\right] -\sqrt{\eta_k}
  tr(\boldsymbol{\Theta}_{(k)\setminus (k)} \boldsymbol{X}_{\setminus (k)} \boldsymbol{\boldsymbol{Y}}^T_{(k)}) \right\},
\end{align*}
which is the claim of (\ref{eq:clusterdraw}).  Note that when $(k)$ is
a singleton, we get the conditional distribution for the alternative
model.

\end{document}